\documentclass[%
 reprint,
superscriptaddress,
nofootinbib,
 amsmath,amssymb,
 aps,
 prd,
]{revtex4-2}

\usepackage{graphicx}
\usepackage{dcolumn}
\usepackage{bm}
\usepackage{xcolor}
\usepackage{orcidlink}

\DeclareRobustCommand{\VAN}[3]{#2}
\let\VANthebibliography\thebibliography
\def\thebibliography{\DeclareRobustCommand{\VAN}[3]{##3}\VANthebibliography}

\newcommand{\msun}{{\,\rm M_\odot}}

\newcommand{\Gyr}{\,{\rm Gyr}}

\newcommand{\pc}{\,{\rm pc}}
\newcommand{\kpc}{\,{\rm kpc}}

\newcommand{\cpm}{\,{\rm cm}^2\,{\rm g}^{-1}}

\newcommand{\kev}{\,{\rm keV}}

\def\jcap{J. Cosmol.  Astropart. Phys.}
\def\aap{A\&A}
\def\apj{ApJ}

\def\apjl{ApJ}
\def\mnras{MNRAS}

\def\na{New Astronomy}

\def\apjs{ApJS}
\def\prd{Phys. Rev. D}

\begin{document}

\preprint{APS/123-QED}

\title{Role of prompt cusps in driving the core collapse of self-interacting dark matter halos}

\author{Vinh Tran\,\orcidlink{0009-0003-6068-6921}}
 \affiliation{Department of Physics and Kavli Institute for Astrophysics and Space Research, Massachusetts Institute of Technology, Cambridge, MA 02139, USA}
 \email{vinhtran@mit.edu}
 
\author{Daniel Gilman\,\orcidlink{0000-0002-5116-7287}}
\affiliation{Department of Astronomy \& Astrophysics, University of Chicago, Chicago, IL 60637, USA}
\affiliation{Brinson Prize Fellow}

\author{M. Sten Delos\,\orcidlink{0000-0003-3808-5321}}
\affiliation{Carnegie Observatories, 813 Santa Barbara Street, Pasadena, CA 91101, USA}

\author{Xuejian Shen\,\orcidlink{0000-0002-6196-823X}}
\affiliation{Department of Physics and Kavli Institute for Astrophysics and Space Research, Massachusetts Institute of Technology, Cambridge, MA 02139, USA}

\author{Oliver Zier\,\orcidlink{0000-0003-1811-8915}}
\affiliation{Center for Astrophysics, Harvard \& Smithsonian, 60 Garden St, Cambridge, MA 02138, USA}

\author{Mark Vogelsberger\,\orcidlink{0000-0001-8593-7692}}
\affiliation{Department of Physics and Kavli Institute for Astrophysics and Space Research, Massachusetts Institute of Technology, Cambridge, MA 02139, USA}

\author{David Xu\,\orcidlink{0009-0007-6324-0131}}
\affiliation{Rice University, Houston, TX 77005, USA}

\date{\today}

\begin{abstract}
    Prompt cusps (PCs) form from the direct collapse of overdensities in the early Universe, reside at the center of every dark matter halo, and have density profiles steeper than $r^{-1}$ NFW cusps. Using a suite of high-resolution N-body simulations, we study the evolution of isolated halos in self-interacting dark matter (SIDM) with massive PCs embedded at their centers, a scenario that could be realized in certain SIDM models with light mediators that predict a small-scale suppression of the linear matter power spectrum. We track the evolution of three equally concentrated $10^7\msun$ halos, hosting PCs of various total masses, and quantify how the presence of a PC affects the processes of core formation and collapse. Early in the core-formation phase, halos with more prominent PCs exhibit a delayed evolution by a factor of $\sim 2$ due to smaller velocity dispersion gradients in the inner region. During most of the core-collapse phase, the halo evolution becomes closely aligned in physical time, with appropriate rescaling of densities, radii, and velocity dispersions. The scale densities and radii preserve the virial mass of the original halos, but with increased concentration. Deviations occur at the late phase of core-collapse at the level of $\sim 5\%$ relative to the reference collapse track of an NFW halo. These deviations depend non-trivially on both the increased concentration incurred by the PCs, as well as the velocity dispersion (temperature) of the outer halo regions, which can inhibit or enhance the heat transfer process. Our simulations illustrate the complex interplay between the inner and outer halo profiles in determining the onset of core collapse and motivate future studies in the full cosmological context.
\end{abstract}

\maketitle


\section{Introduction}
\label{sec:intro}
The internal structures of dark matter (DM) halos encode a wealth of information regarding the particle nature of DM, and the initial conditions for structure formation seeded in the early Universe. For example, self-interacting dark matter (SIDM) \citep{Spergel2000} is a class of well-motivated alternatives to the collisionless cold dark matter (CDM). DM self-interactions conduct heat throughout the halo profile. These theories predict the initial formation of cores inside DM halos, followed by a period of runaway core collapse \citep{Balberg2002,Koda2011}. In the case of the early Universe, recent work has established that primordial density fluctuations do not immediately collapse into structures resembling the universal Navarro-Frenk-White (hereafter NFW) profiles predicted in CDM \citep{Navarro1997}. Instead, primordial density fluctuations collapse into structures referred to as prompt cusps (PCs) \citep{Delos++19,Delos++23,DelosWhite2023,Ondaro-Mallea2024}. 

Most analyses examine the implications of core collapse in the context of classical dwarf galaxies \citep{Nishikawa+20,Correa2021,Nadler+23,Dutra++25}. The impact of SIDM on halo density profiles, particularly the onset of core collapse, is typically regarded as a late-time phenomenon, although some analyses have suggested core collapse as a mechanism to produce early massive black holes \citep{Xiao2021,ShenT2025,Jiang++25,GrantRoberts++25}. Conversely, PCs form from the direct collapse of primordial density fluctuations, and depend on the initial conditions for structure formation and the very early Universe. Despite the different scales and conditions relevant for core collapse and the formation of PCs, these phenomena are connected. The mass of a PC is determined by the cutoff scale of the matter power spectrum, with more-massive PCs forming in models with larger free-streaming lengths or Jeans masses \citep{Delos2023,Delos25}. Some classes of SIDM predict a small-scale suppression of the linear matter power spectrum on the scales relevant for small-scale galactic structure. Therefore, in these classes of SIDM, we expect a relatively massive (up to $\sim 10 \%$ of the total halo mass) PC embedded at the center of every halo. 

The structural evolution of an SIDM halo depends on how quickly heat is conducted from the inner to the outer profile. Injecting a massive, dense structure at the center of a halo (such as a PC), where scattering occurs most frequently, could conceivably impact heat transfer from the inner to the outer halo, delaying or accelerating the onset of core formation and eventual collapse. A delay or acceleration of core collapse in halos could have consequences for observables in this class of theory. Small-scale probes of DM substructure, such as dwarf galaxies \citep{Correa2021,Zeng++25}, stellar streams \citep{Banik++21,Nibauer++25}, and strong gravitational lenses \citep{Gilman+21,Gilman+23,Enzi++25,Powell++25}, have begun to measure halo properties on sub-galactic scales, and could be sensitive to these effects. 

There exist only a limited number of examples of cosmological simulations with SIDM and a suppression of the linear matter power spectrum \citep{Vogelsberger16,Nadler++24}. Recently, \citep{Nadler++24} showed that SIDM models with suppressed matter power spectra exhibit a delay in core collapse. However, these cosmological simulations lack the spatial resolution to resolve PCs inside halos, and do not capture the effect of these structures in affecting the late-time evolution of SIDM halos. In particular, how or whether PCs affect the onset of core collapse remains an open question. Full cosmological simulations with a small-scale cutoff in the matter power spectrum, without enough spatial resolution to resolve PCs, would address this issue. However, this task would push the limits of current computational facilities and codes. 

In this work, we take a first step towards exploring this topic by running a suite of ultra-high-resolution simulations of isolated halos with PCs of various masses embedded at their centers. We initialize the halos using the cusp-halo relation presented by \citet{Delos25}, assuming the initial formation of the prompt cusp in the early Universe is unaffected by self-interactions. We then examine the evolution of these idealized halos in a controlled setting to investigate how a modification to the halo density profile alters the processes of core formation and collapse. We evolve halos with a velocity-dependent self-interaction cross section corresponding to an SIDM model that also predicts a suppression of the linear matter power spectrum \citep{Nadler++24}. We track these halos until the onset of core collapse, and compare their evolution with an NFW profile evolving with the same SIDM cross section to assess the impact of the PC on core formation and collapse. These simulations build on previous high-resolution simulations of SIDM halos designed to explore the phenomenon of core collapse with additional complexity in the SIDM model and halo density profile \citep{Tran2024_1,Tran2025_2}. 

This paper is organized as follows: Section~\ref{sec:sim} presents the simulation setup, including the details of the SIDM and PC models, the initial halo configurations, and the N-body simulation parameters. Section~\ref{sec:halo_evolution} describes the results of the simulations, focusing on the time evolution of the halo cores and comparisons between halos with PCs and the reference NFW halo. Finally, Section~\ref{sec:conclusion} summarizes our main findings and discusses the implications for the evolution of halos hosting prompt cusps.

\section{Simulation Setup}
\label{sec:sim}

\begin{table*}
    \centering
    \addtolength{\tabcolsep}{6.75pt}
    \def\arraystretch{1.8}
    \begin{tabular}{c c c c c c c c c c c}
        \hline
        Halo & $m_{200}$ & $N_{\rm{part}}$ & $N_{200}$ & $m_{\rm DM}$ & $r_{200}$ & $r_{\rm{s}}$ & $\log \rho_{\rm{s}}$ & $y$ & $\epsilon$ & $\eta$ \\ [0ex] %
        & [${\rm M}_\odot$] &  &  & [${\rm M}_\odot$] & [$\rm kpc$]  & [$\rm kpc$] & [${\rm M}_\odot\,{\rm kpc}^{-3}$] &  & [$\rm{pc}$] &  \\ [1ex] %
        \hline\hline

        SIDM-control & $1 \!\times\! 10^7$ & $3.79 \!\times\! 10^6$ & $2 \!\times\! 10^6$ & 5 & 1.667 & 0.352 & 7.297 & 0.000 & 3.1 & 0.0015 \\

        SIDM-low & $1 \!\times\! 10^7$ & $3.81 \!\times\! 10^6$ & $2 \!\times\! 10^6$ & 5 & 1.667 & 0.373 & 7.229 & 0.242 & 3.1 & 0.0015 \\

        SIDM-high & $1 \!\times\! 10^7$ & $3.89 \!\times\! 10^6$ & $2 \!\times\! 10^6$ & 5 & 1.667 & 0.454 & 7.003 & 0.513 & 3.1 & 0.0015 \\

        SIDM-extreme & $1 \!\times\! 10^7$ & $4.01 \!\times\! 10^6$ & $2 \!\times\! 10^6$ & 5 & 1.667 & 0.704 & 6.480 & 1.000 & 3.1 & 0.0015 \\

        \hline
    \end{tabular}
    \caption{Simulations configurations. Each of the listed halos is evolved with DM self-interaction under a velocity-dependent cross section (VDCS) following Equation \ref{eqn:cross_section} and a velocity-independent cross section (VICS) of $\sigma/m = 83.86\cpm$. (2) $M_{200}$ and (6) $r_{200}$ are the virial mass and radius of the halo. (4) $N_{200}$ is the numbers of DM particles within $r_{200}$. (5) $m_{\rm DM}$ is the mass of DM particles. (7) $r_{\rm{s}}$, (8) $\rho_{\rm{s}}$, and (9) $y$ are the parameters for the parametric density profile detailed in Equation \ref{eqn:prompt-cusp_profile}. (10) $\epsilon$ is the (Plummer-equivalent) gravitational softening length of DM particles, and (11) $\eta$ controls the particle gravitational acceleration-based timestep.}
    \label{tab:runs_config}
\end{table*}

This section describes the N-body simulations we run to examine the effects of a prompt cusp on SIDM halo evolution. Section \ref{ssec:modelsetup} describes the DM model that predicts self-interactions alongside a suppression of small-scale structure, as well as the self-interaction cross section used to evolve the halos. The initial halo profiles, including the model for the PCs and the NFW envelopes in which they reside, are also detailed. Section \ref{ssec:profilesetup} details the numerical setup of the N-body simulations.

\subsection{Prompt cusp density profiles and SIDM model}
\label{ssec:modelsetup}

\begin{figure}
    \centering
    \includegraphics[width= 0.49 \textwidth]{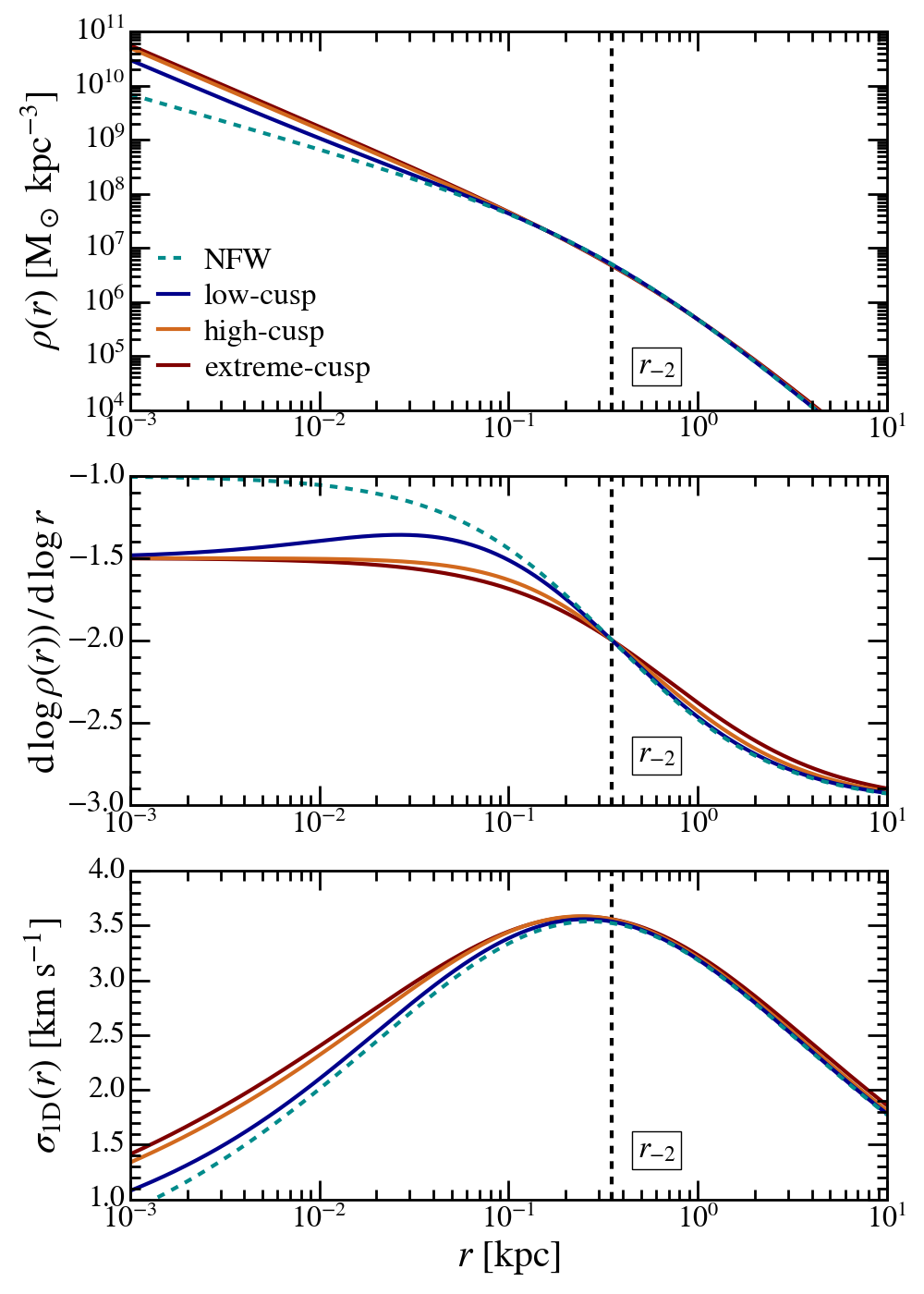}
    \caption{The density profiles (top), logarithmic density profile slopes (middle), and one-dimensional velocity dispersion profiles (bottom) for the four halos described in Section~\ref{ssec:modelsetup}. The reference NFW profile is shown in dashed cyan, while the low-cusp, high-cusp, and extreme-cusp configurations are shown in blue, orange, and red, respectively. The vertical dashed lines represent $r_{-2}$, the radius at which the logarithmic slope reaches the value of $-2$, which coincides with the scale radius $r_{\rm{s}}$ of the reference NFW profile.}
    \label{fig:initial_profiles}
\end{figure}

Building on \citep{Delos++23}, we consider the formation of PC in the early Universe in the context of SIDM. Although \citep{Delos++23} did not explicitly consider the effects of SIDM on PC formation, in order to study the late-time impact of PC on halo evolution, we assume that the formation process proceeds in much the same way as it would in warm dark matter (WDM). This is a reasonable assumption as in the case of WDM, the initial collapses of PCs begin around $z \sim 10 \text{--} 20$, after which it takes roughly $25 \%$ of the Hubble time for them to fully form \cite{Delos++23}. At $z = 10$, this corresponds to $\sim 0.2 \Gyr$, which is much shorter than the timescale we later consider for self-interactions of DM. Therefore, we believe there is no significant difference in the PC formation process between WDM and SIDM with small-scale structure suppression.

We simulate halos that could have formed in a universe with the particle physics model used by~\citep{Nadler++24} with a kinematic decoupling temperature $T_{\rm{kd}}=1.46 \kev$. This model exhibits a cutoff in the matter power spectrum comparable to that of a $6.5 \kev$ thermal relic WDM model, with a suppression of small-scale structure that manifests on scales between $10^7 \text{--} 10^{8}\msun$.
We aim to understand how a prompt cusp at the center of a low-mass halo impacts the time evolution of the halo profile. For what follows, we assume a halo of virial mass $M_{200} = 10^7\msun$ with a concentration $c = 4.74$ at redshift $z=3$, comparable to some of the lowest-mass structures expected to form for the corresponding matter power spectrum. The concentration is defined to be $c=r_{200}/r_{-2}$, where $r_{200}$ is the virial radius and $r_{-2}$ is the radius where the logarithmic density profile slope crosses $-2$. We note that here, the assumption that self-interactions do not have a significant impact on the halo evolution prior to $z = 3$ is adopted. This may not be entirely correct, as seen later in the evolution of the DM halos, and represents a common limitation of SIDM simulations that start from NFW-like ICs.
Using the model presented by ~\citep{Delos25}, we inject prompt cusps into the centers of these halos. The composite PC+NFW halo profile is given by \cite{Delos25}
\begin{equation}
    \label{eqn:prompt-cusp_profile}
    \rho (r \!\mid\! r_{\rm{s}}, \rho_{\rm{s}}, y) = \rho_{\rm{s}} \frac{\sqrt{y^2 + r/r_{\rm{s}}}}{\left(r/r_{\rm{s}}\right)^{3/2}\left(1 + r/r_{\rm{s}}\right)^2}
\end{equation}
where $y = A \, / \rho_{\rm{s}} \, r_{\rm{s}}^{1.5}$. $\rho_{\rm{s}}$ and $r_{\rm{s}}$ are density and radial normalization parameters (which are set by the mass and concentration \cite{Delos25}).
The parameter $A$ sets the amplitude of a prompt cusp profile $\rho(r) \propto A r^{-1.5}$. 

We simulate four halos with properties selected according to
the cusp-halo relation presented by \citet{Delos25}: 
\begin{enumerate}
    \item {\bf{Halo 1 -- SIDM-control}}: This has a NFW profile $\left(y=0\right)$ with $r_{\rm{s}} = 0.352 \kpc$ and $\rho_{\rm{s}} = 1.983\!\times\!10^7 \msun \kpc^{-3}$, serving as a point of reference to be compared with the halos having prompt cusps.
    \item {\bf{Halo 2 -- SIDM-low}}: This halo contains a PC with amplitude $A$ 30\% below the median for this halo mass.
    The halo has parameters $\left(r_{\rm{s}}, \rho_{\rm{s}}, y\right) = \left(0.373 \kpc, 1.694\!\times\!10^7 \msun \kpc^{-3}, 0.242\right)$. These parameters correspond to a typical initial cusp mass of about $3.6 \%$ of the total halo mass.
    \item {\bf{Halo 3 -- SIDM-high}}: This halo contains a PC with amplitude $A$ 30\% above the median, such that the initial cusp mass is about $4.7 \%$ of the total halo mass. It has parameters $\left(r_{\rm{s}}, \rho_{\rm{s}}, y\right) = \left(0.454 \kpc, 1.007\!\times\!10^7 \msun \kpc^{-3}, 0.513\right)$.
    \item {\bf{Halo 4 -- SIDM-extreme}}: This halo contains a PC with amplitude $A$ 47\% above the median, corresponding to an initial cusp mass of $5.0 \%$ of the total halo mass.
    It has parameters $\left(r_{\rm{s}}, \rho_{\rm{s}}, y\right) = \left(0.704 \kpc, 0.302\!\times\!10^7 \msun \kpc^{-3}, 1.000\right)$. With $y=1$, this halo maximizes the influence of the PC given the same fixed concentration parameter $c=4.74$.
\end{enumerate}
The density profiles of these halos are shown in Figure \ref{fig:initial_profiles}. 

We assume an isotropic SIDM cross section
\begin{equation}
    \label{eqn:cross_section}
    \sigma(v) =  \sigma_0 \left(1+v^2/w^2\right)^{-2}
\end{equation}
with $\sigma_0 / m = 147 \ \rm{cm^2} \rm{g^{-1}}$ and $w=24 \ \rm{km} \ \rm{s^{-1}}$. This cross section corresponds to the $T_{\rm{kd}}=1.46 \rm{keV}$ SIDM model, which exhibits a small-scale cutoff in the matter power spectrum expected to give rise to halos comparable to the three cases described above. The values of this cross section place the halos in the long mean-free-path (LMFP) regime throughout their evolution. This parametric model for the scattering cross section closely matches the cross section that results from a Yukawa potential in the Born approximation. In the case of DM halos evolved with self-interactions from NFW ICs, the evolution can be approximated using an effective cross section and collapse timescale. We examine these scale parameters to determine whether the same description remains valid for halos embedded with PCs. Following~\citep{Yang+2022,Yang+23,Tran2025_2}, we approximate halos evolved under such a velocity-dependent cross section (VDCS) by an effective velocity-independent cross section (VICS) with a strength given by \citep{Yang+2022,Yang+23}
\begin{equation}
    \label{eqn:sigma_kappa}
    \sigma_{\rm{eff}} = \sigma_{\kappa} \equiv\frac{\langle \sigma (v) \, v^5 \rangle}{\langle v^5 \rangle}.
\end{equation}
Here, the brackets signify the expected values assuming the isotropic Maxwell-Boltzmann (MB) velocity distribution, i.e.,
\begin{equation}
    \label{eqn:MB_averaging}
    \langle \mathcal{O} (v) \rangle \equiv \frac{1}{2\sqrt{\pi} \, \sigma_{\rm{1D}}^3} \int_0^\infty \mathcal{O} (v) \, {\rm{d}} v \, v^2 \exp{\left(- \frac{v^2}{4 \sigma_{\rm{1D}}^2}\right)}.
\end{equation}
For halos initialized with the NFW configuration, the effective one-dimensional velocity dispersion $\sigma_{\rm{1D}}$ can be approximated as $\sigma_{\rm{1D}} \approx 1.10 \, \sqrt{G \rho_{\rm{s}} r_{\rm{s}}^2}$ following~\citep{Tran2025_2}, with $G$ here being the gravitational constant. Using the parameters of the SIDM-control halo, the effective cross section comes out to be $\sigma_{\kappa} = 83.86\cpm$. 

We will compare the time evolution of halo profiles in terms of a characteristic collapse timescale \citep{Essig2019,Yang2024}
\begin{equation}
    \label{eqn:collapse_timescale}
    \tau(\sigma_{\rm{eff}}/m) = \frac{150}{C} \frac{1}{\sigma_{\rm eff}/m} \frac{1}{\rho_{\rm{eff}}} \left(\frac{1}{4 \pi \, G \, \rho_{\rm{eff}} \, r_{\rm{eff}}^2}\right)^{1/2},
\end{equation}
where $\sigma_{\rm eff} = 83.86\cpm$ is the effective cross section. $\rho_{\rm{eff}}$, and $r_{\rm{eff}}$ represent some choice of characteristic scale density and scale radius, chosen here to be those of the reference NFW profile (i.e. Halo 1). $C$ is taken to be $C \simeq 0.85$ following~\citep{Tran2025_1} to approximate $T = \tau$ as the onset of the gravothermal catastrophe, when the core density increases exponentially.

\subsection{N-body simulation details}
\label{ssec:profilesetup}

\begin{figure}
    \centering
    \includegraphics[width= 0.49 \textwidth]{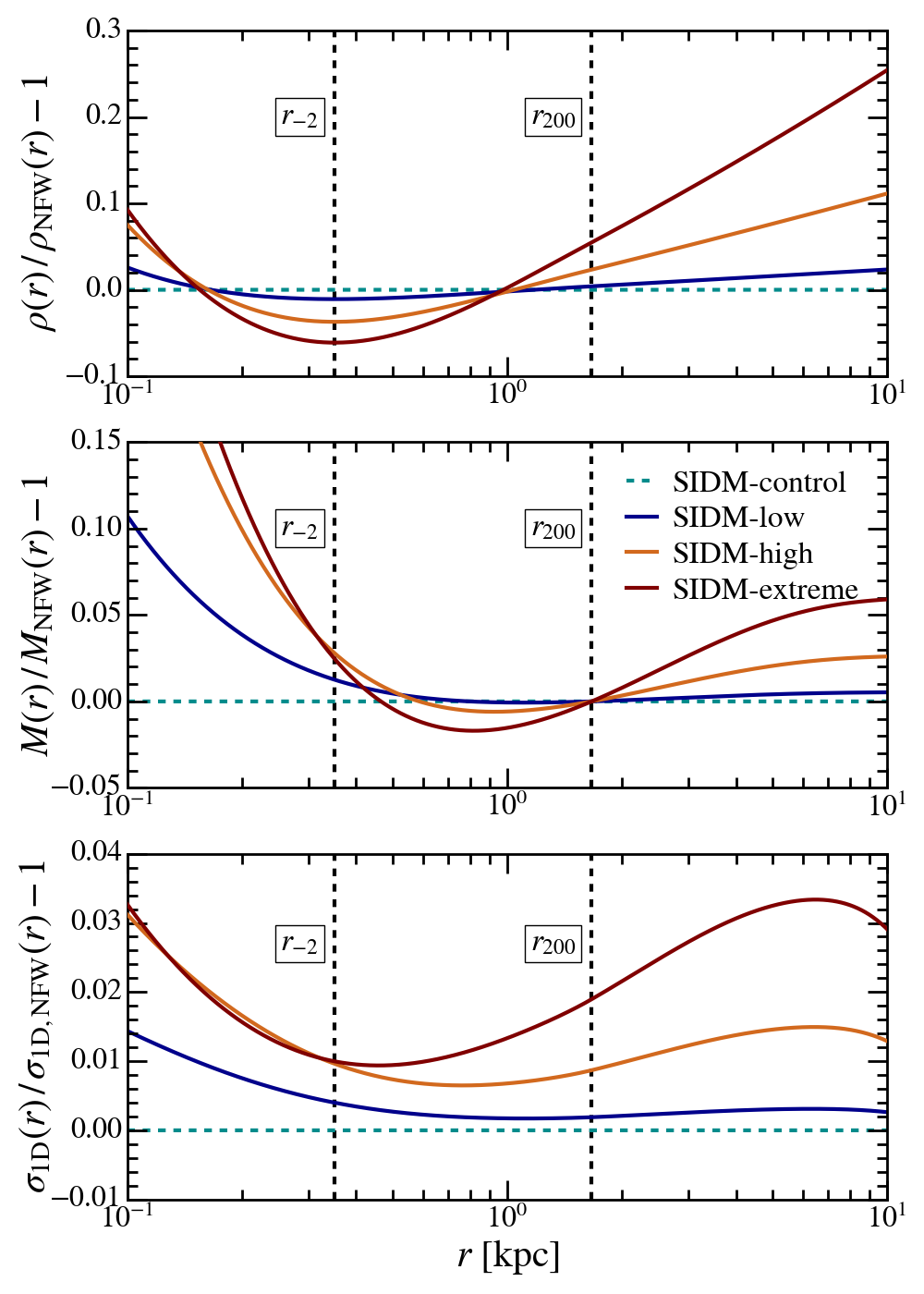}
    \caption{The fractional differences in the initial density profiles $\rho(r)$ (top), enclosed mass profiles $M(r)$ (middle), and one-dimensional velocity dispersion $\sigma_{\rm{1D}} (r)$ (bottom) of the four halos relative to the reference NFW halo (SIDM-control). $r_{-2}$ and $r_{200}$ are indicated by vertical dashed lines. As a result of the functional form of the parametric model and the exponential cut-off model beyond $r_{200}$, the halos differ not only in the prompt cusp–dominated region but also across the entire halo, particularly in the outer regions. These differences may influence the subsequent time evolution of the halo core during the core-collapse phase.}
    \label{fig:mass_ratio}
\end{figure}

We perform dark-matter-only N-body simulations of idealized isolated halos using the multi-physics, massively parallel simulation code \textsc{Arepo} \cite{Springel2010,Weinberger2020}. For all simulations, gravity is computed using the Tree method, with the softening length as suggested in \cite{Mace+24}. The corresponding Plummer-equivalent softening length adopted in our runs is $\epsilon = 3.1\pc$. DM self-interactions are modeled following the implementation in \cite{Vogelsberger2012}. The gravitational timesteps of the particles are scaled by the parameter $\eta$ as ${\rm{d}}t = \sqrt{2 \eta \epsilon / \left| a \right|}$. Here, $\epsilon$ is the aforementioned softening length, and $\left| a \right|$ is the particle acceleration. Based on the 
results of~\citep{Mace+24} and the approximated collapse timescale of $\tau \sim 43.13\Gyr$, we choose $\eta = 0.0015$. 

All halos are initialized using stable configurations derived in the CDM model, with adjustments made to account for the softened gravitational field, as detailed in Appendix \ref{apd:ICs}. The initial conditions (ICs) code \textsc{SoftIsoICs} is publicly available\footnote{\url{https://github.com/vinh-qtran/SoftIsoICs}}. The ICs of our halos are generated following the parametric density profile detailed in Equation \ref{eqn:prompt-cusp_profile} within the virial radius ($r_{200}$) and a modified exponential cut-off outside $r_{200}$~\citep{Springel1999,Kazantzidis2004}. The profiles take the analytical forms of
\begin{align}
    \label{eqn:inner_density}
    \rho_{r \leq r_{200}} (r) &= \rho (r \!\mid\! r_{\rm{s}}, \rho_{\rm{s}}, y) \\
    \label{eqn:outer_density}
    \rho_{r > r_{200}} (r) &= \rho_{200}\left(\frac{r}{r_{200}}\right)^{\epsilon_{\rm d}} \exp\left(-\frac{r-r_{200}}{r_{\rm d}}\right),
\end{align}
with $\rho_{200} = \rho (r_{200} \!\mid\! r_{\rm{s}}, \rho_{\rm{s}}, y)$. $r_{\rm d}$ is the decay scale radius, taken to be $r_{200}$ for simplicity, while $\epsilon_{\rm d}$ is the exponential decay index, chosen so that the continuity of the logarithmic slope of the density profile is preserved
\begin{equation}
    \label{eqn:nfw_decay_exponential}
    \epsilon_{\rm d} = \frac{r_{200}}{r_{\rm d}} + \left. \frac{{\rm d} \log \rho}{{\rm d} \log r} \right|_{r_{200}}.
\end{equation}
This exponential cut-off is designed to model the physical truncation of DM halo in the cosmological context. 

Figure~\ref{fig:mass_ratio} shows the initial density, enclosed mass, and one-dimensional velocity dispersion ratios between the prompt-cusp halos and control NFW configurations. Each prompt-cusp halo exhibits a higher central density than the NFW profile, owing to the presence of the PC, and also an enhanced density beyond $r_{200}$. In this outer region, the density profile of the extreme-cusp halo deviates from those of the low-cusp and control halos by $10\text{--}15\%$, and from the high-cusp halo by $\sim 5\%$, corresponding to enclosed-mass deviations of $\sim 5\%$ and $\sim 2\%$, respectively. These differences arise directly from the chosen parameterization of the halo density profile, as well as from the exponential cutoff applied to the outer regions. The increased mass at large radii also produces a corresponding increase in the velocity dispersion in those regions. Although these deviations are relatively small, they may nonetheless affect the subsequent collapse times by altering the temperature gradient, and consequently the efficiency of heat transport.

For each halo detailed in \ref{ssec:modelsetup}, we perform one simulation using the DM self-interacting VICS of $\sigma / m = \sigma_\kappa = 83.86\cpm$ and another with the VDCS described by Equation \ref{eqn:cross_section}. Each halo is initialized out to $5 \, r_{200}$ and contains $N \sim 4 \times 10^6$ particles, approximately half of which residing within $r_{200}$. The inner region of each halo (defined here as $r \leq 0.1 \, r_{-2}$) is resolved with at least $1 \!\times\! 10^4$ particles. This numerical resolution ensures we meet convergence criteria for simulating gravothermal collapse \citep{Mace+24}. To ensure that the structural evolutions observed in the simulations are physical rather than numerical artifacts, we also perform the same set of halo simulations initialized using different random number generator seeds, along with auxiliary runs conducted at varying resolutions and timestep criteria. Table \ref{tab:runs_config} summarizes the main simulation configuration.

\section{Time-evolution of SIDM halos with prompt cusps}
\label{sec:halo_evolution}

\subsection{The general evolution of halo structures}
\label{ssec:structure}

\begin{figure}
    \centering
    \includegraphics[width= 0.49 \textwidth]{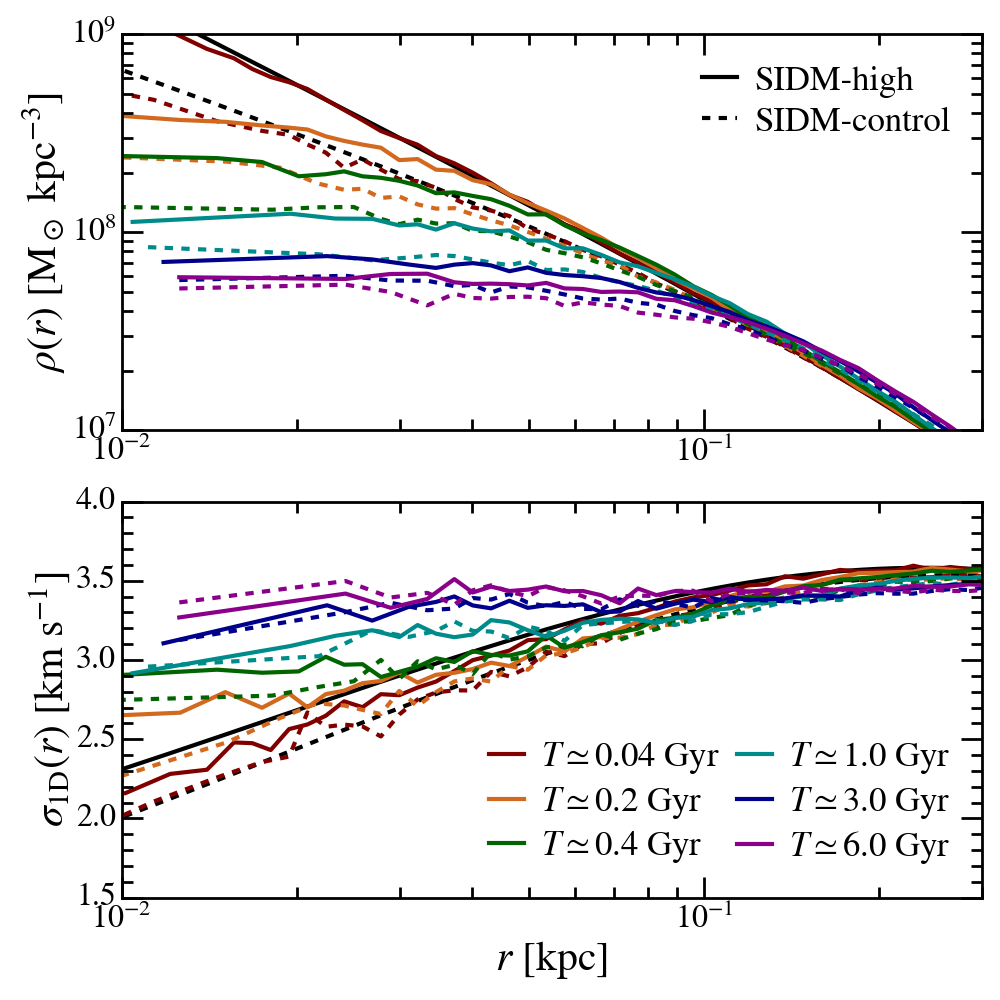}
    \caption{The evolution of the density (top) and velocity dispersion (bottom) profiles of the high-cusp (solid) and control NFW (dashed) halos during the core-formation phase. Several snapshots are chosen, spanning from the earliest epoch, approximately $0.1\%$ of the collapse timescale, to $T \sim 6\Gyr$, which marks the transition into the core-collapse phase. Overall, the high-cusp halo evolves significantly slower than the NFW reference until the onset of core collapse, at which point the two halos begin to exhibit similar internal structures.}
    \label{fig:init_evolution}
\end{figure}

\begin{figure}
    \centering
    \includegraphics[width= 0.49 \textwidth]{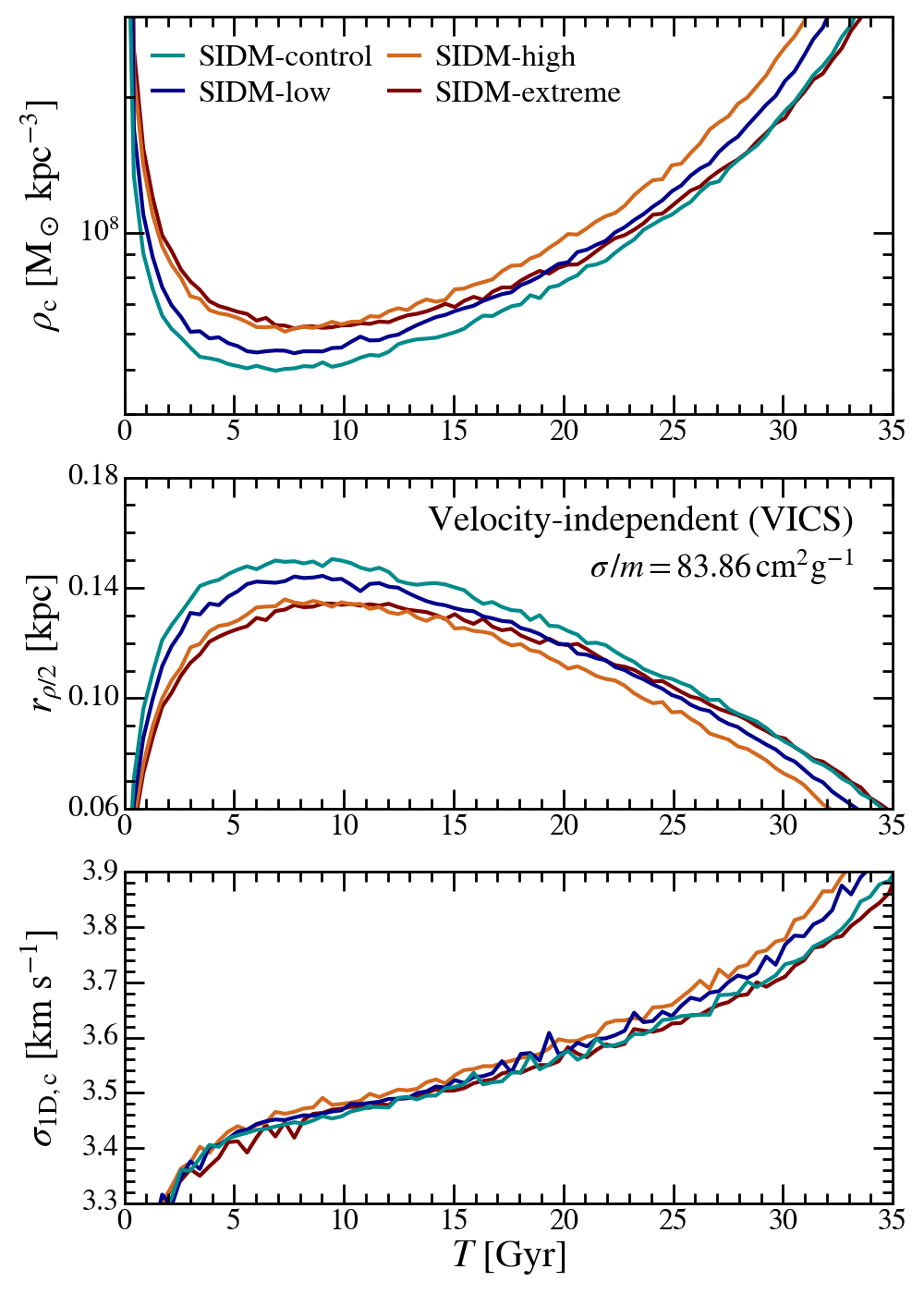}
    \caption{Evolutions of the core density $\rho_{\rm{c}}$ (top), core half-density radius $r_{\rho/2}$ (middle), and core one-dimensional velocity dispersion $\sigma_{\rm{1D,c}}$ (bottom) of the four halos in runs with the VICS per particle mass of $\sigma / m = 83.86\cpm$. As a result of their initially denser cusps, the prompt-cusp halos reach higher minimum core densities and smaller half-density radii compared to the NFW reference.}
    \label{fig:core_evolution}
\end{figure}

\begin{figure*}
    \centering
    \includegraphics[width= 0.99 \textwidth]{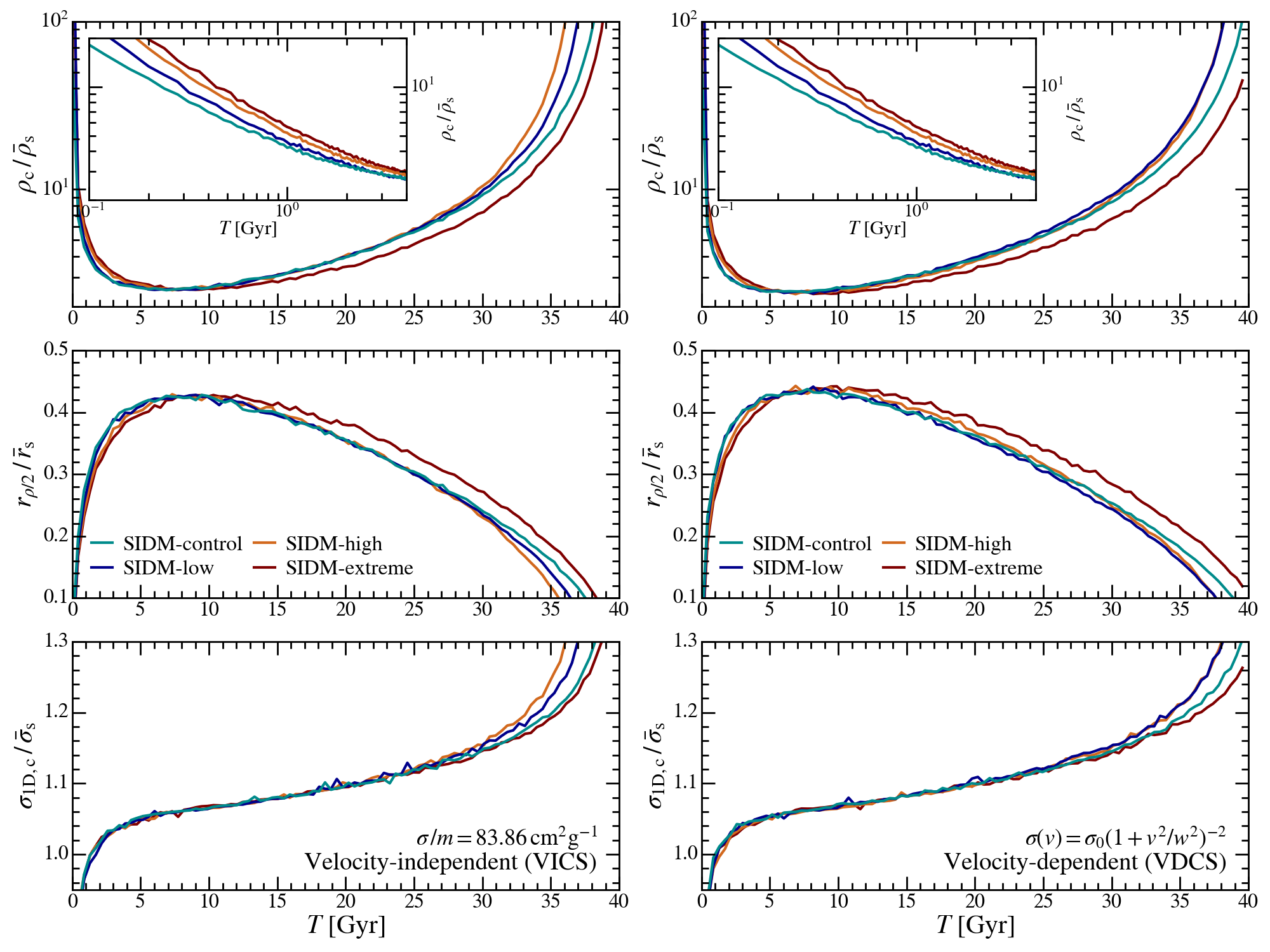}
    \caption{Scaled evolutions of the core density $\rho_{\rm{c}}$ (top), core half-density radius $r_{\rho/2}$ (middle), and core one-dimensional velocity dispersion $\sigma_{\rm{1D,c}}$ (bottom) for the four halos in VDCS (left) and VICS (right) models. The sub-panels show the zoomed-in scaled core density evolution during the core-formation phase. The effective scale density $\bar{\rho}_{\rm{s}}$, effective scale radius $\bar{r}_{\rm{s}}$, and effective scale velocity $\sigma_{\rm{s}}$ are taken as shown in Table \ref{tab:collapse_params}. Generally, the evolutionary trajectories of the different halos appear broadly overlapping, with the exception of the extreme-cusp halo. The prompt-cusp halos are also observed to form cores more gradually while undergoing core collapse at accelerated rates.}
    \label{fig:core_evolution_scaled}
\end{figure*}

\begin{table*}
    \centering
    \addtolength{\tabcolsep}{6.25pt}
    \def\arraystretch{1.8}
    \begin{tabular}{c c c c c c c}
        \hline
        Halo & $\bar{\rho}_{\rm{s}}$ & $\bar{r}_{\rm{s}}$ & $\bar{\sigma}_{\rm{s}}$ & $\sqrt{G \bar{\rho}_{\rm{s}} \bar{r}_{\rm{s}}^2}$ & $\bar{m}_{200}$ & $\bar{c}$ \\ [0ex] %
        & [$\times 10^7 \, {\rm M}_\odot\,{\rm kpc}^{-3}$] & [$\rm kpc$] & [${\rm{km}} \, {\rm{s}}^{-1}$] & [${\rm{km}} \, {\rm{s}}^{-1}$] & [$\times 10^7 \, {\rm M}_\odot$] &  \\ [1ex] %
        \hline\hline

        SIDM-control & $1.983$ & $0.352$ & $3.249$ & $3.249$ & $1.000$ & $4.739$ \\

        SIDM-low & $2.162 \pm 0.014$ & $0.338 \pm 0.002$ & $3.262 \pm 0.007$ & $3.244 \pm 0.024$ & $0.994 \pm 0.015$ & $4.936 \pm 0.029$ \\

        SIDM-high & $2.422 \pm 0.016$ & $0.320 \pm 0.002$ & $3.271 \pm 0.007$ & $3.250 \pm 0.023$ & $0.984 \pm 0.015$ & $5.211 \pm 0.032$ \\

        SIDM-extreme & $2.453 \pm 0.016$ & $0.313 \pm 0.002$ & $3.251 \pm 0.006$ & $3.218 \pm 0.021$ & $0.950 \pm 0.014$ & $5.322 \pm 0.031$ \\

        \hline
    \end{tabular}
    \caption{The effective scale parameters for each halo. The effective scale density (1) $\bar{\rho}_{\rm{s}}$ and effective radius (2) $\bar{r}_{\rm{s}}$ are calculated from the minimum core density and maximum core half-density radius in VICS runs, while the effective scale velocity (3) $\bar{\sigma}_{\rm{s}}$ are calculated from the slope of the core one-dimensional velocity dispersion. (4) $\bar{\sigma}_{\rm{s}}^\prime = \sqrt{G \bar{\rho}_{\rm{s}} \bar{r}_{\rm{s}}^2}$ represent the values of $\bar{\sigma}_{\rm{s}}$ calculated from $\bar{\rho}_{\rm{s}}$ and $\bar{r}_{\rm{s}}$. The resulting effective NFW virial mass (5) $\bar{m}_{200}$ and concentration parameters (6) $\bar{c}$ are also shown. The more pronounced the prompt cusp, the more concentrated the corresponding NFW-equivalent halo. However, although the virial masses are in good agreement with the expected values, the effective velocities deviate from the anticipated trends of increasing velocity dispersion.}
    \label{tab:collapse_params}
\end{table*}

Similar to what has been observed in other studies of elastic DM self-interactions~\citep[e.g.][]{Vogelsberger2012,Yang+23,Tran2025_2}, our halos undergo an initial core-formation phase, characterized by the development of an isothermal, constant-density core, followed by gravothermal core collapse~\citep{LyndenBell68}. Figure \ref{fig:init_evolution} shows the evolution of the high-cusp and control NFW halos during the core-formation phase. Due to being initially much denser, owing to the prompt cusp, the high-cusp halo develops a core at a noticeably slower pace than the NFW reference. This trend is evident in both the density and velocity dispersion profiles. Nevertheless, after $T \sim 3\Gyr$, the two halos begin to display comparable internal kinetic structures, even though their density profiles still differ by several tens of percent in the central regions.
To quantify the evolution of SIDM halos more precisely, we characterize the SIDM halo cores using the core density ($\rho_{\rm{c}}$), the core half-density radius ($r_{\rho/2}$, defined as the radius at which the density drops to half the value at the core), and the core one-dimensional velocity dispersion ($\sigma_{\rm{1D,c}}$), following the procedures detailed in \cite{Tran2025_2}. Specifically, $\sigma_{\rm{1D,c}}$ is measured directly from the one-dimensional velocity dispersion profiles of the halos using an iterative hypothesis testing procedure (following Section III.C of~\citep{Tran2025_2}). $\rho_{\rm{c}}$ and $r_{\rho/2}$ are obtained by fitting the density profile using the function form
\begin{equation}
    \label{eqn:TVS_density_profile}
    \rho_{\rm{T24}} (r) = \rho_{\rm{c}} \left( \frac{\tanh{r/r_{\rm{c}}}}{r/r_{\rm{c}}} \right)^n \frac{1}{\left(1 + \left(r/r_{\rm{s}}^{\prime}\right)^2\right)^{\left(3-n\right)/2}}.
\end{equation}
Here, $r_{\rm c}$ is the characteristic core radius, from which the core half-density radius can be calculated following $r_{\rho/2} = r_{\rm{c}} \,f^{-1}(0.5^{1/n})$, where $f^{-1}$ is the inverse function of $f(x) \equiv \tanh{x}/x$. $r_{\rm{s}}^{\prime}$ is a scale radius similar to (but not to be confused with) the NFW scale radius $r_{\rm{s}}$. $n$ controls the steepness of the transition from the constant-density core ($\rho \sim \text{const}$) and the NFW tail ($\rho \sim r^{-3}$). As demonstrated in~\citep{Tran2025_1}, the profile has the advantage of closely approximating both the flat-core (i.e. $\rho \sim \text{const}$) and isothermal-core (i.e. $\sigma_{\rm{1D}} \sim \text{const}$) configurations in the inner regions, serving as an accurate parametric form for the halo density profile. The uncertainties of the fitted parameters are calculated following a bootstrapping approach, that is, perturbing the density profile according to the measured uncertainties, repeating the fitting process for each sample, and calculating the standard deviations of the resulting parameters.

Figure \ref{fig:core_evolution} shows the time evolution of the core density $\rho_{\rm{c}}$, core half-density radius $r_{\rho/2}$, and core one-dimensional velocity dispersion $\sigma_{\rm{1D,c}}$ of the four halos in the VICS runs. The prompt-cusp halos reach higher core densities and smaller core sizes compared to the reference NFW halo. The magnitude of these deviations correlates with the prominence of the initial PCs. Based on the central velocity dispersion, the prompt-cusp halos develop slightly hotter cores, likely a consequence of their initially elevated central temperatures (as shown in Figure \ref{fig:initial_profiles}). Interestingly, despite the high-cusp and extreme-cusp halos having similar initial conditions, their evolution trajectories in the core-collapse phase diverge significantly, the reason for which will be examined in greater detail in subsequent parts of this paper. 

Despite the differences in their absolute values, the evolutionary tracks of the various halos follow similar trajectories. This motivates us to identify characteristic scaling factors for the density, core size, and velocity dispersion. We compute these values using the core characteristics at the epoch of minimum core density and maximum core radius, which occur at approximately the same time for all halos. This epoch corresponds to the transition from the core-formation phase to the core-collapse regime, and, owing to the slow evolution of the system at this stage, provides a stable point at which these quantities can be measured. The exception is the scaling factor for the velocity dispersion, which we discuss further below. In detail, we compute the effective scale densities $\bar{\rho}_{\rm{s}}$, scale radii $\bar{r}_{\rm{s}}$, and scale velocity dispersions $\bar{\sigma}_{\rm{s}}$ as follows
\begin{enumerate}
    \item $\bar{\rho}_{\rm{s}}$ - obtained by scaling the control NFW scale density by the ratio of the minimum core density reached by the prompt-cusp halos (denoted $\rho_{\rm{c}}^{\rm{PC}}$) to that of the NFW control halo (denoted $\rho_{\rm{c}}^{\rm{NFW}}$), i.e. 
    \begin{equation}
        \label{eqn:effective_scale_density}
        \bar{\rho}_{\rm{s}} = \rho_s \frac{\rm{min} (\rho_{\rm{c}}^{\rm{PC}}(T))}{\rm{min} (\rho_{\rm{c}}^{\rm{NFW}}(T))}.
    \end{equation}
    \item $\bar{r}_{\rm{s}}$ - determined by applying the ratios of the maximum core half-density radii to the control NFW scale radius, i.e.
    \begin{equation}
        \label{eqn:effective_scale_radius}
        \bar{r}_{\rm{s}}=r_s \frac{\rm{max} (r_{\rho/2}^{\rm{PC}}(T))}{\rm{max} (r_{\rho/2}^{\rm{NFW}}(T))}.
    \end{equation}
    \item $\bar{\sigma}_{\rm{s}}$ - calculated from the ratios of the rates of increase in the core one-dimensional velocity dispersion over the interval $T = 10 \text{--} 20\Gyr$ ($\sim 0.25 \text{--} 0.5 \, \tau$), during which the velocity dispersion evolves approximately linearly, i.e. 
    \begin{equation}
        \label{eqn:effective_scale_velocity_dispersion}
        \bar{\sigma}_{\rm{s}} = \sigma_{\rm{s}} \frac{\langle \dot{\sigma}_{\rm{1D,c}}^{\rm{PC}}(T=0.25 \text{--} 0.5 \, \tau) \rangle}{\langle \dot{\sigma}_{\rm{1D,c}}^{\rm{NFW}}(T=0.25 \text{--} 0.5 \, \tau) \rangle}
    \end{equation}
    where $\dot{\sigma}$ denotes a time derivative. 
\end{enumerate}
The values are listed in Table~\ref{tab:collapse_params}. We observe that the NFW halos, constructed with the resulting effective scale density and scale radius, preserve the virial radius ($r_{200} = 1.667\kpc$) and virial mass ($m_{200} \simeq 10^7\msun$) of the original halo. We also find that these scaled NFW halos exhibit the same enclosed mass at $r \simeq 0.65\,r_{-2}$ as the original halo with PC profiles (see Appendix \ref{apd:eff_params} for more details). One can therefore predict these effective scale values directly from the ICs.

We observe that when the quantities are normalized by the effective scale density $\bar{\rho}_{\rm{s}}$, scale radius $\bar{r}_{\rm{s}}$, and scale velocity dispersion $\bar{\sigma}_{\rm{s}}$, the time evolutions of the core density, core half-density radius, and core one-dimensional velocity dispersion evolve more similarly over time. Figure~\ref{fig:core_evolution_scaled} shows these scaled evolutions as functions of physical time for the four halos in both the VICS and VDCS runs. Except for the case of the extreme cusp, the core evolutions remain consistent among the halos up to approximately $75\%$ of the gravothermal catastrophe onset. Interestingly, this self-similarity emerges not when scaled by the collapse timescale $\bar{\tau} \propto 1 / \bar{\rho}_{\rm{s}}\bar{\sigma}_{\rm{s}}$, but rather with respect to physical time\footnote{The halos appear to collapse approximately $10\%$ faster than the expected collapse timescale of $\tau = 43.13,\Gyr$, likely as a consequence of outer-halo stripping, as discussed in Appendix~\ref{apd:stripping}.}. A closer examination reveals that the prompt-cusp halos form cores at more gradual rates compared to the reference NFW profile. This rate correlates with the prominence of the prompt cusp, such that halos with stronger cusps develop cores more slowly. These trends are observed in both the VICS and VDCS SIDM models. We discuss this observation further in Section \ref{ssec:core-formation}. After $T \sim 30\Gyr$, the core evolutions of the halos begin to diverge at the level of $5 \text{--} 10 \%$, with the patterns of deviation differing between the VDCS and VICS runs. We attribute these discrepancies to the influence of the outer regions of the halo in absorbing heat from the center, a topic that we discussed in greater detail in Section \ref{ssec:core-collapse}.

\subsection{The core-formation phase \& the erasure of the prompt cusps}
\label{ssec:core-formation}

\begin{figure}
    \centering
    \includegraphics[width= 0.49 \textwidth]{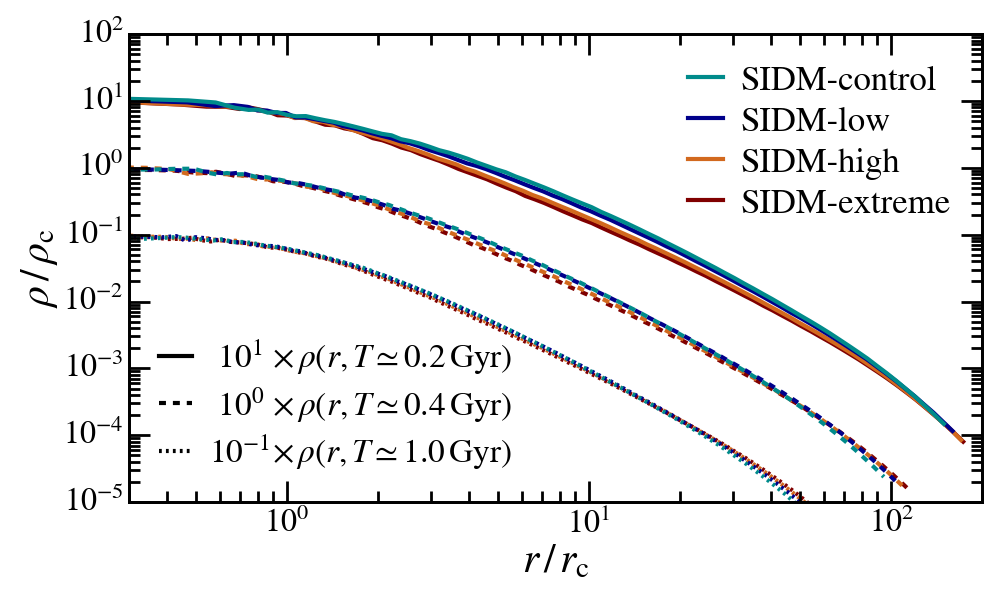}
    \caption{The self-similar density profiles of the four halos in the VICS runs during the earliest epochs of core-formation, at $T \simeq 0.2\Gyr$ (solid), $T \simeq 0.4\Gyr$ (dashed),  and $T \simeq 1.0\Gyr$ (dotted). For visual clarity, the densities of halos at these epochs are scaled by a factor of $10$, $1$, and $0.1$, respectively. At the onset of the core-formation phase, halos containing more pronounced prompt cusps exhibit noticeably steeper slopes outside the core region. However, shortly afterward, all halos converge toward a common self-similar solution, as all traces of the prompt cusps in the density structures vanish.}
    \label{fig:core-formation}
\end{figure}

Figure \ref{fig:core-formation} shows the density profiles of the four halos during the early stages of the core-formation phase. Here, densities and radii are scaled by the core density $\rho_{\rm c}$ and core radius $r_{\rm c}$ to illustrate the degree of self-similarity between halos. We observe that up to $0.4\Gyr$, which corresponds to roughly $1\%$ of the collapse timescale, halos initially embedded with more prominent prompt cusps exhibit steeper density slopes outside the core. This behavior is consistent with the initial conditions shown in Figure \ref{fig:initial_profiles}. As the halo cores evolve and thermalization propagates outward, the halos gradually become more self-similar, and the signatures of the prompt cusps are progressively erased. This erasure is essentially complete by $1\Gyr$, i.e. at approximately $2.5\%$ of the collapse timescale. Nevertheless, as seen in Figure \ref{fig:core_evolution_scaled}, the influence of the initial prompt cusps is not entirely negligible, as halos with larger initial prompt cusps tend to evolve more slowly during the core-formation phase. This effect is most evident at early times, where the core-formation rate can differ by a factor of $\sim 2$ between the extreme-cusp halo and the NFW reference. This can be attributed to the PCs increasing the velocity dispersion (and thus temperature) gradient, which delays the inward heat flow during the core-formation phase. As a result, halos with higher initial prompt cusps, despite having denser cores and higher collision rates, form cores at a substantially slower pace. Once the core transitions into the core-collapse phase, however, the evolutionary tracks of the halos become more closely aligned.


\subsection{Late-time evolution \& core-collapse time}
\label{ssec:core-collapse}

\begin{figure}
    \centering
    \includegraphics[width= 0.49 \textwidth]{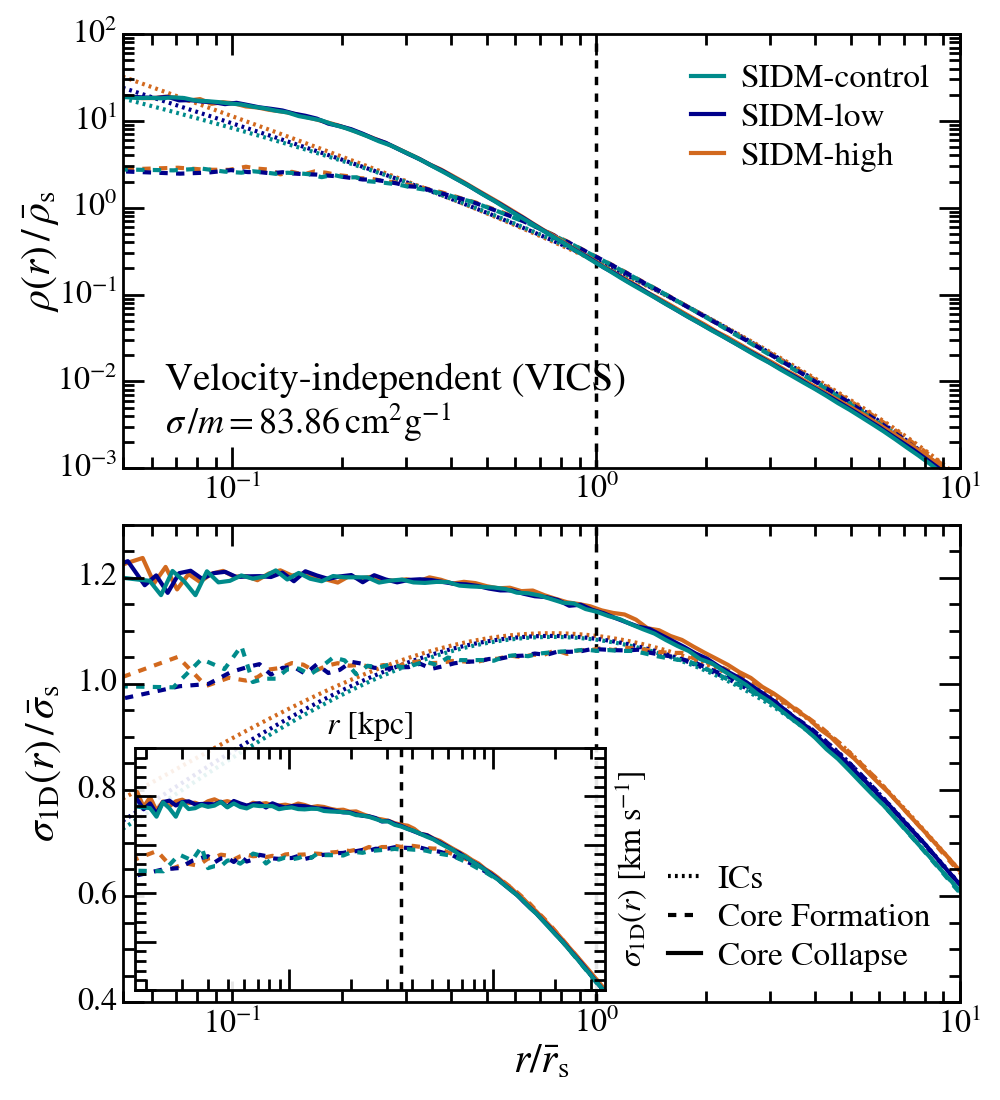}
    \caption{The scaled density profiles (top) and scaled velocity dispersion profiles (bottom) of the control, low-, and high-cusp halos in the VICS model at two representative epochs - one during the core-formation phase (dashed) and another during the core-collapse phase (solid). The ICs are shown as references in dotted lines. The sub-panel presents the corresponding velocity dispersion profiles in physical units. The vertical dashed lines represent $r / \bar{r}_{\rm{s}} = 1$. The core-formation halos are chosen such that $\rho_{\rm{c}} / \bar{\rho}_{\rm{s}} \simeq 2.75$ and $r_{\rho/2} / \bar{r}_{\rm{s}} \simeq 0.393$, while the core-collapse halos satisfying $\rho_{\rm{c}} / \bar{\rho}_{\rm{s}} \simeq 20.5$ and $r_{\rho/2} / \bar{r}_{\rm{s}} \simeq 0.167$. When scaled by the effective scale parameters in Table \ref{tab:collapse_params}, the halos' density profiles appear self-similar. This is, however, not the case for the velocity dispersion profiles, which appear to be more self-similar in physical units, or with only the velocity dispersions scaled by $\sigma_{\rm{s}}$.}
    \label{fig:profiles}
\end{figure}

\begin{table}
    \centering
    \addtolength{\tabcolsep}{11pt}
    \def\arraystretch{1.8}
    \begin{tabular}{c c c c}
        \hline
        Halo & \multicolumn{2}{c}{$\tau_{\rm 100}/\tau_{\rm 100}^{\rm NFW}$} & $\bar{\tau}/\bar{\tau}^{\rm NFW}$\\
         & VICS & VDCS & \\ [1ex]
        \hline\hline
        
        SIDM-low & 0.967 & 0.966 & 0.915 \\
        
        SIDM-high & 0.943 & 0.964 & 0.813 \\
        
        SIDM-extreme & 1.024 & 1.021 & 0.808 \\
        
        \hline
    \end{tabular}
    \caption{The relative collapse times of the prompt-cusp halos with respect to the reference NFW halo. The empirical collapse time $\tau_{\rm 100}$ is defined as the epoch at which the core density reaches $100$ times the effective scale density $\bar{\rho}_{\rm{s}}$. The ratios of the effective collapse timescale $\bar{\tau} \propto 1 / \bar{\rho}_{\rm{s}} \bar{\sigma}_{\rm{s}}$ are also shown. Such ratios, however, do not appear to serve as reliable predictors of the collapse time in prompt cusp halos.}
    \label{tab:collapse_times}  
\end{table}

In Figure~\ref{fig:profiles}, we assess the degree of self-similarity among the halos by examining the alignment of the density and one-dimensional velocity dispersion profiles between the control NFW halo and the high- and low-cusp halos in the VICS model. Snapshots corresponding to the core-formation and core-collapse phases are selected based on their scaled core density $\rho_{\rm{c}} / \bar{\rho}_{\rm{s}}$ and core half-density radius $r_{\rho/2} / \bar{r}_{\rm{s}}$, which typically coincide when the scaled densities are the same between halos. The core-formation and core-collapse snapshots are chosen such that $\rho_{\rm{c}} / \bar{\rho}_{\rm{s}} \simeq 2.75$, $r_{\rho/2} / \bar{r}_{\rm{s}} \simeq 0.393$, and $\rho_{\rm{c}} / \bar{\rho}_{\rm{s}} \simeq 20.5$, $r_{\rho/2} / \bar{r}_{\rm{s}} \simeq 0.167$, respectively. We find that, when scaled by $\bar{\rho}_{\rm{s}}$ and $\bar{r}_{\rm{s}}$, the density profiles of the inner regions exhibit strong self-similarity among the halos. This suggests that the specific structural imprints introduced by the prompt cusps are effectively erased once the density profiles are properly scaled. In contrast, the one-dimensional velocity dispersion profiles, when scaled by $\bar{\sigma}_{\rm{s}}$ and $\bar{r}_{\rm{s}}$, show deviations of a few percent beyond $\bar{r}_{\rm{s}}$. In the unscaled profiles, however, the velocity dispersions become closely aligned. 

Table~\ref{tab:collapse_times} presents the core collapse times of the prompt-cusp halos relative to the control NFW halos evolving under the same cross sections. The collapse time is defined as the epoch when the core density reaches $100$ times the effective scale density. The effective collapse timescales, $\sim 1 / \bar{\rho}_{\rm{s}}\bar{\sigma}_{\rm{s}}$ tend to deviate at the level of $20 \%$. For the low- and high-cusp halos, the core collapse proceeds more rapidly, correlating with the prominences of the initial PCs, at the later stage of evolution. In contrast, the extreme-cusp halo exhibits a delayed core collapse. 


As discussed in Section \ref{ssec:profilesetup}, excess mass residing beyond $r_{200}$ observed in Figure \ref{fig:mass_ratio} can plausibly decelerate the collapse time of halos: this additional material increases the temperature in the outskirts, decreasing the gradient and suppressing the outward heat transport, which slows down the core collapse. We see this trend for NFW profiles as well, shown in Figure~\ref{fig:extended_evolution} (see Appendix \ref{apd:stripping}), where the collapse of the control NFW halo is accelerated by the deficit of material in the outer regions compared to a more extended configuration. However, not every halo with excess material in the outskirts exhibits decelerated core collapse; the low and high-cusp halos both collapse slightly faster than the NFW control. On the other hand, the extreme-cusp halo collapses more slowly. 

From these observations, we infer that there may be multiple competing factors that govern the overall collapse timescale. Among them is the increase in central density, which acts to accelerate core collapse. The second likely contributing factor is the influence of the outer boundary, which can either impede or enhance the collapse depending on the thermal state of the outer region. The same observations made in this section for halos evolved under the VICS model also apply to halos evolved under VDCS (see Appendix \ref{apd:vdcs_profiles}). The slight deviations between the two cases can be attributed to the additional complexity in the collapse times introduced by the velocity dependence.

\section{Discussion and conclusions}
\label{sec:conclusion}

In this work, we perform a series of $N$-body simulations of idealized, isolated DM halos with embedded prompt cusps with properties based on theoretical expectations. The prompt cusps are introduced into NFW profiles using a parametric density model that preserves both the virial mass $m_{200}$ and radius $r_{200}$, as well as the radius at which the logarithmic density slope equals $-2$, denoted as $r_{-2}$. We compare the time evolution of the core density, core half-density radius, and core one-dimensional velocity dispersion of the prompt-cusp halos in both the VICS and VDCS models with that of the control NFW halo. Our main findings are as follows:
\begin{enumerate}
    \item DM self-interactions rapidly thermalize the halos, erasing the signatures of the PCs in both the density and velocity dispersion profiles. However, because the PCs initially incur a lower temperature gradient, halos with more prominent PCs form cores more slowly.
    \item We find that halos with embedded PCs follow a nearly self-similar evolution once they are rescaled to characteristic densities, radii, and velocity dispersions. These values deviate at a level of $\sim 10\%$ among halos with PCs and the NFW reference. 
    \item During the late core-collapse phase, interplay between increasing central density and the elevated velocity dispersion in the outer halo exert competing influences on the collapse time.
\end{enumerate}

In general, the most significant impact of the PC is the delay introduced in core formation. Although brief, during the first $\sim 0.04\Gyr$ (approximately $1\%$ of the collapse timescale), halos initially embedded with more prominent PCs also take longer to develop self-similar density profiles. Once the core has formed and the system enters the core-collapse phase, however, the evolution becomes nearly self-similar in both VICS and VDCS runs, provided that the corresponding adjustments to the scale density and scale radius are taken into account. The inferred increase in concentration appears to be predictable, i.e. it can be obtained by determining the concentration for which the enclosed mass at $0.65\,r_{-2}$ of an NFW profile with the same virial mass matches that of the initial prompt-cusp profile. The alignment of halo evolution during the core-collapse phase occurs in physical time rather than in units of the effective timescale, suggesting that the overall collapse time of a halo may depend on more complex factors and combinations of quantities, rather than solely on the local collision timescale.


Once DM self-interactions are activated, the cores of the halos rapidly thermalize and become approximately self-similar, erasing the characteristic $\rho\left(r\right) \propto r^{-1.5}$ structure of the prompt cusps. It remains a question whether prompt cusps would survive the initial collapse of a primordial density fluctuation when DM self-interactions are in place and persist at the center of a halo in the first place. Cosmological simulations, or idealized simulations of spherical collapse with self-interactions, would help address this open question.

\begin{acknowledgments}
We thank Francis-Yan Cyr-Racine, Hai-bo Yu, and Moritz Fischer for useful discussions. We thank the anonymous referee for useful comments and suggestions. The simulations in this paper were conducted on the Engaging cluster at Massachusetts Institute of Technology (MIT) operated by the MIT Office of Research Computing and Data. DG acknowledges support for this work provided by the Brinson Foundation through a Brinson Prize Fellowship grant. XS acknowledges the support from the National Aeronautics and Space Administration (NASA) theory grant JWST-AR-04814. Support for OZ was provided by Harvard University through the Institute for Theory and Computation Fellowship.
\end{acknowledgments}

\bibliography{main}

\providecommand{\noopsort}[1]{}\providecommand{\singleletter}[1]{#1}%
\begin{thebibliography}{47}%
\makeatletter
\providecommand \@ifxundefined [1]{%
 \@ifx{#1\undefined}
}%
\providecommand \@ifnum [1]{%
 \ifnum #1\expandafter \@firstoftwo
 \else \expandafter \@secondoftwo
 \fi
}%
\providecommand \@ifx [1]{%
 \ifx #1\expandafter \@firstoftwo
 \else \expandafter \@secondoftwo
 \fi
}%
\providecommand \natexlab [1]{#1}%
\providecommand \enquote  [1]{``#1''}%
\providecommand \bibnamefont  [1]{#1}%
\providecommand \bibfnamefont [1]{#1}%
\providecommand \citenamefont [1]{#1}%
\providecommand \href@noop [0]{\@secondoftwo}%
\providecommand \href [0]{\begingroup \@sanitize@url \@href}%
\providecommand \@href[1]{\@@startlink{#1}\@@href}%
\providecommand \@@href[1]{\endgroup#1\@@endlink}%
\providecommand \@sanitize@url [0]{\catcode `\\12\catcode `\$12\catcode `\&12\catcode `\#12\catcode `\^12\catcode `\_12\catcode `\%12\relax}%
\providecommand \@@startlink[1]{}%
\providecommand \@@endlink[0]{}%
\providecommand \url  [0]{\begingroup\@sanitize@url \@url }%
\providecommand \@url [1]{\endgroup\@href {#1}{\urlprefix }}%
\providecommand \urlprefix  [0]{URL }%
\providecommand \Eprint [0]{\href }%
\providecommand \doibase [0]{https://doi.org/}%
\providecommand \selectlanguage [0]{\@gobble}%
\providecommand \bibinfo  [0]{\@secondoftwo}%
\providecommand \bibfield  [0]{\@secondoftwo}%
\providecommand \translation [1]{[#1]}%
\providecommand \BibitemOpen [0]{}%
\providecommand \bibitemStop [0]{}%
\providecommand \bibitemNoStop [0]{.\EOS\space}%
\providecommand \EOS [0]{\spacefactor3000\relax}%
\providecommand \BibitemShut  [1]{\csname bibitem#1\endcsname}%
\let\auto@bib@innerbib\@empty
\bibitem [{\citenamefont {{Spergel}}\ and\ \citenamefont {{Steinhardt}}(2000)}]{Spergel2000}%
  \BibitemOpen
  \bibfield  {author} {\bibinfo {author} {\bibfnamefont {D.~N.}\ \bibnamefont {{Spergel}}}\ and\ \bibinfo {author} {\bibfnamefont {P.~J.}\ \bibnamefont {{Steinhardt}}},\ }\bibfield  {title} {\bibinfo {title} {{Observational Evidence for Self-Interacting Cold Dark Matter}},\ }\href {https://doi.org/10.1103/PhysRevLett.84.3760} {\bibfield  {journal} {\bibinfo  {journal} {\prl}\ }\textbf {\bibinfo {volume} {84}},\ \bibinfo {pages} {3760} (\bibinfo {year} {2000})},\ \Eprint {https://arxiv.org/abs/astro-ph/9909386} {arXiv:astro-ph/9909386 [astro-ph]} \BibitemShut {NoStop}%
\bibitem [{\citenamefont {{Balberg}}\ \emph {et~al.}(2002)\citenamefont {{Balberg}}, \citenamefont {{Shapiro}},\ and\ \citenamefont {{Inagaki}}}]{Balberg2002}%
  \BibitemOpen
  \bibfield  {author} {\bibinfo {author} {\bibfnamefont {S.}~\bibnamefont {{Balberg}}}, \bibinfo {author} {\bibfnamefont {S.~L.}\ \bibnamefont {{Shapiro}}},\ and\ \bibinfo {author} {\bibfnamefont {S.}~\bibnamefont {{Inagaki}}},\ }\bibfield  {title} {\bibinfo {title} {{Self-Interacting Dark Matter Halos and the Gravothermal Catastrophe}},\ }\href {https://doi.org/10.1086/339038} {\bibfield  {journal} {\bibinfo  {journal} {\apj}\ }\textbf {\bibinfo {volume} {568}},\ \bibinfo {pages} {475} (\bibinfo {year} {2002})},\ \Eprint {https://arxiv.org/abs/astro-ph/0110561} {arXiv:astro-ph/0110561 [astro-ph]} \BibitemShut {NoStop}%
\bibitem [{\citenamefont {{Koda}}\ and\ \citenamefont {{Shapiro}}(2011)}]{Koda2011}%
  \BibitemOpen
  \bibfield  {author} {\bibinfo {author} {\bibfnamefont {J.}~\bibnamefont {{Koda}}}\ and\ \bibinfo {author} {\bibfnamefont {P.~R.}\ \bibnamefont {{Shapiro}}},\ }\bibfield  {title} {\bibinfo {title} {{Gravothermal collapse of isolated self-interacting dark matter haloes: N-body simulation versus the fluid model}},\ }\href {https://doi.org/10.1111/j.1365-2966.2011.18684.x} {\bibfield  {journal} {\bibinfo  {journal} {\mnras}\ }\textbf {\bibinfo {volume} {415}},\ \bibinfo {pages} {1125} (\bibinfo {year} {2011})},\ \Eprint {https://arxiv.org/abs/1101.3097} {arXiv:1101.3097 [astro-ph.CO]} \BibitemShut {NoStop}%
\bibitem [{\citenamefont {{Navarro}}\ \emph {et~al.}(1997)\citenamefont {{Navarro}}, \citenamefont {{Frenk}},\ and\ \citenamefont {{White}}}]{Navarro1997}%
  \BibitemOpen
  \bibfield  {author} {\bibinfo {author} {\bibfnamefont {J.~F.}\ \bibnamefont {{Navarro}}}, \bibinfo {author} {\bibfnamefont {C.~S.}\ \bibnamefont {{Frenk}}},\ and\ \bibinfo {author} {\bibfnamefont {S.~D.~M.}\ \bibnamefont {{White}}},\ }\bibfield  {title} {\bibinfo {title} {{A Universal Density Profile from Hierarchical Clustering}},\ }\href {https://doi.org/10.1086/304888} {\bibfield  {journal} {\bibinfo  {journal} {\apj}\ }\textbf {\bibinfo {volume} {490}},\ \bibinfo {pages} {493} (\bibinfo {year} {1997})},\ \Eprint {https://arxiv.org/abs/astro-ph/9611107} {arXiv:astro-ph/9611107 [astro-ph]} \BibitemShut {NoStop}%
\bibitem [{\citenamefont {{Delos}}\ \emph {et~al.}(2019)\citenamefont {{Delos}}, \citenamefont {{Bruff}},\ and\ \citenamefont {{Erickcek}}}]{Delos++19}%
  \BibitemOpen
  \bibfield  {author} {\bibinfo {author} {\bibfnamefont {M.~S.}\ \bibnamefont {{Delos}}}, \bibinfo {author} {\bibfnamefont {M.}~\bibnamefont {{Bruff}}},\ and\ \bibinfo {author} {\bibfnamefont {A.~L.}\ \bibnamefont {{Erickcek}}},\ }\bibfield  {title} {\bibinfo {title} {{Predicting the density profiles of the first halos}},\ }\href {https://doi.org/10.1103/PhysRevD.100.023523} {\bibfield  {journal} {\bibinfo  {journal} {\prd}\ }\textbf {\bibinfo {volume} {100}},\ \bibinfo {eid} {023523} (\bibinfo {year} {2019})},\ \Eprint {https://arxiv.org/abs/1905.05766} {arXiv:1905.05766 [astro-ph.CO]} \BibitemShut {NoStop}%
\bibitem [{\citenamefont {{Delos}}\ and\ \citenamefont {{White}}(2023{\natexlab{a}})}]{Delos++23}%
  \BibitemOpen
  \bibfield  {author} {\bibinfo {author} {\bibfnamefont {M.~S.}\ \bibnamefont {{Delos}}}\ and\ \bibinfo {author} {\bibfnamefont {S.~D.~M.}\ \bibnamefont {{White}}},\ }\bibfield  {title} {\bibinfo {title} {{Inner cusps of the first dark matter haloes: formation and survival in a cosmological context}},\ }\href {https://doi.org/10.1093/mnras/stac3373} {\bibfield  {journal} {\bibinfo  {journal} {\mnras}\ }\textbf {\bibinfo {volume} {518}},\ \bibinfo {pages} {3509} (\bibinfo {year} {2023}{\natexlab{a}})},\ \Eprint {https://arxiv.org/abs/2207.05082} {arXiv:2207.05082 [astro-ph.CO]} \BibitemShut {NoStop}%
\bibitem [{\citenamefont {{Delos}}\ and\ \citenamefont {{White}}(2023{\natexlab{b}})}]{DelosWhite2023}%
  \BibitemOpen
  \bibfield  {author} {\bibinfo {author} {\bibfnamefont {M.~S.}\ \bibnamefont {{Delos}}}\ and\ \bibinfo {author} {\bibfnamefont {S.~D.~M.}\ \bibnamefont {{White}}},\ }\bibfield  {title} {\bibinfo {title} {{Prompt cusps and the dark matter annihilation signal}},\ }\href {https://doi.org/10.1088/1475-7516/2023/10/008} {\bibfield  {journal} {\bibinfo  {journal} {\jcap}\ }\textbf {\bibinfo {volume} {2023}},\ \bibinfo {eid} {008} (\bibinfo {year} {2023}{\natexlab{b}})},\ \Eprint {https://arxiv.org/abs/2209.11237} {arXiv:2209.11237 [astro-ph.CO]} \BibitemShut {NoStop}%
\bibitem [{\citenamefont {{Ondaro-Mallea}}\ \emph {et~al.}(2024)\citenamefont {{Ondaro-Mallea}}, \citenamefont {{Angulo}}, \citenamefont {{St{\"u}cker}}, \citenamefont {{Hahn}},\ and\ \citenamefont {{White}}}]{Ondaro-Mallea2024}%
  \BibitemOpen
  \bibfield  {author} {\bibinfo {author} {\bibfnamefont {L.}~\bibnamefont {{Ondaro-Mallea}}}, \bibinfo {author} {\bibfnamefont {R.~E.}\ \bibnamefont {{Angulo}}}, \bibinfo {author} {\bibfnamefont {J.}~\bibnamefont {{St{\"u}cker}}}, \bibinfo {author} {\bibfnamefont {O.}~\bibnamefont {{Hahn}}},\ and\ \bibinfo {author} {\bibfnamefont {S.~D.~M.}\ \bibnamefont {{White}}},\ }\bibfield  {title} {\bibinfo {title} {{Phase-space simulations of prompt cusps: simulating the formation of the first haloes without artificial fragmentation}},\ }\href {https://doi.org/10.1093/mnras/stad3949} {\bibfield  {journal} {\bibinfo  {journal} {\mnras}\ }\textbf {\bibinfo {volume} {527}},\ \bibinfo {pages} {10802} (\bibinfo {year} {2024})},\ \Eprint {https://arxiv.org/abs/2309.05707} {arXiv:2309.05707 [astro-ph.GA]} \BibitemShut {NoStop}%
\bibitem [{\citenamefont {{Nishikawa}}\ \emph {et~al.}(2020)\citenamefont {{Nishikawa}}, \citenamefont {{Boddy}},\ and\ \citenamefont {{Kaplinghat}}}]{Nishikawa+20}%
  \BibitemOpen
  \bibfield  {author} {\bibinfo {author} {\bibfnamefont {H.}~\bibnamefont {{Nishikawa}}}, \bibinfo {author} {\bibfnamefont {K.~K.}\ \bibnamefont {{Boddy}}},\ and\ \bibinfo {author} {\bibfnamefont {M.}~\bibnamefont {{Kaplinghat}}},\ }\bibfield  {title} {\bibinfo {title} {{Accelerated core collapse in tidally stripped self-interacting dark matter halos}},\ }\href {https://doi.org/10.1103/PhysRevD.101.063009} {\bibfield  {journal} {\bibinfo  {journal} {\prd}\ }\textbf {\bibinfo {volume} {101}},\ \bibinfo {eid} {063009} (\bibinfo {year} {2020})},\ \Eprint {https://arxiv.org/abs/1901.00499} {arXiv:1901.00499 [astro-ph.GA]} \BibitemShut {NoStop}%
\bibitem [{\citenamefont {{Correa}}(2021)}]{Correa2021}%
  \BibitemOpen
  \bibfield  {author} {\bibinfo {author} {\bibfnamefont {C.~A.}\ \bibnamefont {{Correa}}},\ }\bibfield  {title} {\bibinfo {title} {{Constraining velocity-dependent self-interacting dark matter with the Milky Way's dwarf spheroidal galaxies}},\ }\href {https://doi.org/10.1093/mnras/stab506} {\bibfield  {journal} {\bibinfo  {journal} {\mnras}\ }\textbf {\bibinfo {volume} {503}},\ \bibinfo {pages} {920} (\bibinfo {year} {2021})},\ \Eprint {https://arxiv.org/abs/2007.02958} {arXiv:2007.02958 [astro-ph.GA]} \BibitemShut {NoStop}%
\bibitem [{\citenamefont {{Nadler}}\ \emph {et~al.}(2023)\citenamefont {{Nadler}}, \citenamefont {{Yang}},\ and\ \citenamefont {{Yu}}}]{Nadler+23}%
  \BibitemOpen
  \bibfield  {author} {\bibinfo {author} {\bibfnamefont {E.~O.}\ \bibnamefont {{Nadler}}}, \bibinfo {author} {\bibfnamefont {D.}~\bibnamefont {{Yang}}},\ and\ \bibinfo {author} {\bibfnamefont {H.-B.}\ \bibnamefont {{Yu}}},\ }\bibfield  {title} {\bibinfo {title} {{A Self-interacting Dark Matter Solution to the Extreme Diversity of Low-mass Halo Properties}},\ }\href {https://doi.org/10.3847/2041-8213/ad0e09} {\bibfield  {journal} {\bibinfo  {journal} {\apjl}\ }\textbf {\bibinfo {volume} {958}},\ \bibinfo {eid} {L39} (\bibinfo {year} {2023})},\ \Eprint {https://arxiv.org/abs/2306.01830} {arXiv:2306.01830 [astro-ph.GA]} \BibitemShut {NoStop}%
\bibitem [{\citenamefont {{Dutra}}\ \emph {et~al.}(2025)\citenamefont {{Dutra}}, \citenamefont {{Natarajan}},\ and\ \citenamefont {{Gilman}}}]{Dutra++25}%
  \BibitemOpen
  \bibfield  {author} {\bibinfo {author} {\bibfnamefont {I.}~\bibnamefont {{Dutra}}}, \bibinfo {author} {\bibfnamefont {P.}~\bibnamefont {{Natarajan}}},\ and\ \bibinfo {author} {\bibfnamefont {D.}~\bibnamefont {{Gilman}}},\ }\bibfield  {title} {\bibinfo {title} {{Self-interacting Dark Matter, Core Collapse, and the Galaxy{\textendash}Galaxy Strong-lensing Discrepancy}},\ }\href {https://doi.org/10.3847/1538-4357/ad9b09} {\bibfield  {journal} {\bibinfo  {journal} {\apj}\ }\textbf {\bibinfo {volume} {978}},\ \bibinfo {eid} {38} (\bibinfo {year} {2025})},\ \Eprint {https://arxiv.org/abs/2406.17024} {arXiv:2406.17024 [astro-ph.CO]} \BibitemShut {NoStop}%
\bibitem [{\citenamefont {{Xiao}}\ \emph {et~al.}(2021)\citenamefont {{Xiao}}, \citenamefont {{Shen}}, \citenamefont {{Hopkins}},\ and\ \citenamefont {{Zurek}}}]{Xiao2021}%
  \BibitemOpen
  \bibfield  {author} {\bibinfo {author} {\bibfnamefont {H.}~\bibnamefont {{Xiao}}}, \bibinfo {author} {\bibfnamefont {X.}~\bibnamefont {{Shen}}}, \bibinfo {author} {\bibfnamefont {P.~F.}\ \bibnamefont {{Hopkins}}},\ and\ \bibinfo {author} {\bibfnamefont {K.~M.}\ \bibnamefont {{Zurek}}},\ }\bibfield  {title} {\bibinfo {title} {{SMBH seeds from dissipative dark matter}},\ }\href {https://doi.org/10.1088/1475-7516/2021/07/039} {\bibfield  {journal} {\bibinfo  {journal} {\jcap}\ }\textbf {\bibinfo {volume} {2021}},\ \bibinfo {eid} {039} (\bibinfo {year} {2021})},\ \Eprint {https://arxiv.org/abs/2103.13407} {arXiv:2103.13407 [astro-ph.CO]} \BibitemShut {NoStop}%
\bibitem [{\citenamefont {{Shen}}\ \emph {et~al.}(2025)\citenamefont {{Shen}}, \citenamefont {{Shen}}, \citenamefont {{Xiao}}, \citenamefont {{Vogelsberger}},\ and\ \citenamefont {{Jiang}}}]{ShenT2025}%
  \BibitemOpen
  \bibfield  {author} {\bibinfo {author} {\bibfnamefont {T.}~\bibnamefont {{Shen}}}, \bibinfo {author} {\bibfnamefont {X.}~\bibnamefont {{Shen}}}, \bibinfo {author} {\bibfnamefont {H.}~\bibnamefont {{Xiao}}}, \bibinfo {author} {\bibfnamefont {M.}~\bibnamefont {{Vogelsberger}}},\ and\ \bibinfo {author} {\bibfnamefont {F.}~\bibnamefont {{Jiang}}},\ }\bibfield  {title} {\bibinfo {title} {{Massive Black Holes Seeded by Dark Matter -- Implications for Little Red Dots and Gravitational Wave Signatures}},\ }\href {https://doi.org/10.48550/arXiv.2504.00075} {\bibfield  {journal} {\bibinfo  {journal} {arXiv e-prints}\ ,\ \bibinfo {eid} {arXiv:2504.00075}} (\bibinfo {year} {2025})},\ \Eprint {https://arxiv.org/abs/2504.00075} {arXiv:2504.00075 [astro-ph.GA]} \BibitemShut {NoStop}%
\bibitem [{\citenamefont {{Jiang}}\ \emph {et~al.}(2025)\citenamefont {{Jiang}}, \citenamefont {{Jia}}, \citenamefont {{Zheng}}, \citenamefont {{Ho}}, \citenamefont {{Inayoshi}}, \citenamefont {{Shen}}, \citenamefont {{Vogelsberger}},\ and\ \citenamefont {{Feng}}}]{Jiang++25}%
  \BibitemOpen
  \bibfield  {author} {\bibinfo {author} {\bibfnamefont {F.}~\bibnamefont {{Jiang}}}, \bibinfo {author} {\bibfnamefont {Z.}~\bibnamefont {{Jia}}}, \bibinfo {author} {\bibfnamefont {H.}~\bibnamefont {{Zheng}}}, \bibinfo {author} {\bibfnamefont {L.~C.}\ \bibnamefont {{Ho}}}, \bibinfo {author} {\bibfnamefont {K.}~\bibnamefont {{Inayoshi}}}, \bibinfo {author} {\bibfnamefont {X.}~\bibnamefont {{Shen}}}, \bibinfo {author} {\bibfnamefont {M.}~\bibnamefont {{Vogelsberger}}},\ and\ \bibinfo {author} {\bibfnamefont {W.-X.}\ \bibnamefont {{Feng}}},\ }\bibfield  {title} {\bibinfo {title} {{Formation of the Little Red Dots from the Core-collapse of Self-interacting Dark Matter Halos}},\ }\href {https://doi.org/10.48550/arXiv.2503.23710} {\bibfield  {journal} {\bibinfo  {journal} {arXiv e-prints}\ ,\ \bibinfo {eid} {arXiv:2503.23710}} (\bibinfo {year} {2025})},\ \Eprint {https://arxiv.org/abs/2503.23710} {arXiv:2503.23710 [astro-ph.GA]} \BibitemShut {NoStop}%
\bibitem [{\citenamefont {{Grant Roberts}}\ \emph {et~al.}(2025)\citenamefont {{Grant Roberts}}, \citenamefont {{Braff}}, \citenamefont {{Garg}}, \citenamefont {{Profumo}}, \citenamefont {{Jeltema}},\ and\ \citenamefont {{O'Donnell}}}]{GrantRoberts++25}%
  \BibitemOpen
  \bibfield  {author} {\bibinfo {author} {\bibfnamefont {M.}~\bibnamefont {{Grant Roberts}}}, \bibinfo {author} {\bibfnamefont {L.}~\bibnamefont {{Braff}}}, \bibinfo {author} {\bibfnamefont {A.}~\bibnamefont {{Garg}}}, \bibinfo {author} {\bibfnamefont {S.}~\bibnamefont {{Profumo}}}, \bibinfo {author} {\bibfnamefont {T.}~\bibnamefont {{Jeltema}}},\ and\ \bibinfo {author} {\bibfnamefont {J.}~\bibnamefont {{O'Donnell}}},\ }\bibfield  {title} {\bibinfo {title} {{Early formation of supermassive black holes from the collapse of strongly self-interacting dark matter}},\ }\href {https://doi.org/10.1088/1475-7516/2025/01/060} {\bibfield  {journal} {\bibinfo  {journal} {\jcap}\ }\textbf {\bibinfo {volume} {2025}},\ \bibinfo {eid} {060} (\bibinfo {year} {2025})},\ \Eprint {https://arxiv.org/abs/2410.17480} {arXiv:2410.17480 [astro-ph.GA]} \BibitemShut {NoStop}%
\bibitem [{\citenamefont {{Delos}}(2023)}]{Delos2023}%
  \BibitemOpen
  \bibfield  {author} {\bibinfo {author} {\bibfnamefont {M.~S.}\ \bibnamefont {{Delos}}},\ }\bibfield  {title} {\bibinfo {title} {{Massive prompt cusps: a new signature of warm dark matter}},\ }\href {https://doi.org/10.1093/mnrasl/slad043} {\bibfield  {journal} {\bibinfo  {journal} {\mnras}\ }\textbf {\bibinfo {volume} {522}},\ \bibinfo {pages} {L78} (\bibinfo {year} {2023})},\ \Eprint {https://arxiv.org/abs/2302.03040} {arXiv:2302.03040 [astro-ph.CO]} \BibitemShut {NoStop}%
\bibitem [{\citenamefont {{Delos}}(2025)}]{Delos25}%
  \BibitemOpen
  \bibfield  {author} {\bibinfo {author} {\bibfnamefont {M.~S.}\ \bibnamefont {{Delos}}},\ }\bibfield  {title} {\bibinfo {title} {{The cusp-halo relation}},\ }\href {https://doi.org/10.48550/arXiv.2506.03240} {\bibfield  {journal} {\bibinfo  {journal} {arXiv e-prints}\ ,\ \bibinfo {eid} {arXiv:2506.03240}} (\bibinfo {year} {2025})},\ \Eprint {https://arxiv.org/abs/2506.03240} {arXiv:2506.03240 [astro-ph.CO]} \BibitemShut {NoStop}%
\bibitem [{\citenamefont {{Zeng}}\ \emph {et~al.}(2025)\citenamefont {{Zeng}}, \citenamefont {{Peter}}, \citenamefont {{Du}}, \citenamefont {{Benson}}, \citenamefont {{Li}}, \citenamefont {{Mace}},\ and\ \citenamefont {{Yang}}}]{Zeng++25}%
  \BibitemOpen
  \bibfield  {author} {\bibinfo {author} {\bibfnamefont {Z.~C.}\ \bibnamefont {{Zeng}}}, \bibinfo {author} {\bibfnamefont {A.~H.~G.}\ \bibnamefont {{Peter}}}, \bibinfo {author} {\bibfnamefont {X.}~\bibnamefont {{Du}}}, \bibinfo {author} {\bibfnamefont {A.}~\bibnamefont {{Benson}}}, \bibinfo {author} {\bibfnamefont {J.}~\bibnamefont {{Li}}}, \bibinfo {author} {\bibfnamefont {C.}~\bibnamefont {{Mace}}},\ and\ \bibinfo {author} {\bibfnamefont {S.}~\bibnamefont {{Yang}}},\ }\bibfield  {title} {\bibinfo {title} {{Diversity and universality: Evolution of dwarf galaxies with self-interacting dark matter}},\ }\href {https://doi.org/10.1103/x9t4-3zy7} {\bibfield  {journal} {\bibinfo  {journal} {\prd}\ }\textbf {\bibinfo {volume} {112}},\ \bibinfo {eid} {063008} (\bibinfo {year} {2025})},\ \Eprint {https://arxiv.org/abs/2412.14621} {arXiv:2412.14621 [astro-ph.GA]} \BibitemShut {NoStop}%
\bibitem [{\citenamefont {{Banik}}\ \emph {et~al.}(2021)\citenamefont {{Banik}}, \citenamefont {{Bovy}}, \citenamefont {{Bertone}}, \citenamefont {{Erkal}},\ and\ \citenamefont {{de Boer}}}]{Banik++21}%
  \BibitemOpen
  \bibfield  {author} {\bibinfo {author} {\bibfnamefont {N.}~\bibnamefont {{Banik}}}, \bibinfo {author} {\bibfnamefont {J.}~\bibnamefont {{Bovy}}}, \bibinfo {author} {\bibfnamefont {G.}~\bibnamefont {{Bertone}}}, \bibinfo {author} {\bibfnamefont {D.}~\bibnamefont {{Erkal}}},\ and\ \bibinfo {author} {\bibfnamefont {T.~J.~L.}\ \bibnamefont {{de Boer}}},\ }\bibfield  {title} {\bibinfo {title} {{Novel constraints on the particle nature of dark matter from stellar streams}},\ }\href {https://doi.org/10.1088/1475-7516/2021/10/043} {\bibfield  {journal} {\bibinfo  {journal} {\jcap}\ }\textbf {\bibinfo {volume} {2021}},\ \bibinfo {eid} {043} (\bibinfo {year} {2021})},\ \Eprint {https://arxiv.org/abs/1911.02663} {arXiv:1911.02663 [astro-ph.GA]} \BibitemShut {NoStop}%
\bibitem [{\citenamefont {{Nibauer}}\ \emph {et~al.}(2025)\citenamefont {{Nibauer}}, \citenamefont {{Bonaca}}, \citenamefont {{Price-Whelan}}, \citenamefont {{Spergel}},\ and\ \citenamefont {{Greene}}}]{Nibauer++25}%
  \BibitemOpen
  \bibfield  {author} {\bibinfo {author} {\bibfnamefont {J.}~\bibnamefont {{Nibauer}}}, \bibinfo {author} {\bibfnamefont {A.}~\bibnamefont {{Bonaca}}}, \bibinfo {author} {\bibfnamefont {A.~M.}\ \bibnamefont {{Price-Whelan}}}, \bibinfo {author} {\bibfnamefont {D.~N.}\ \bibnamefont {{Spergel}}},\ and\ \bibinfo {author} {\bibfnamefont {J.~E.}\ \bibnamefont {{Greene}}},\ }\bibfield  {title} {\bibinfo {title} {{Measurement of Dark Matter Substructure from the Kinematics of the GD-1 Stellar Stream}},\ }\href {https://doi.org/10.48550/arXiv.2510.02247} {\bibfield  {journal} {\bibinfo  {journal} {arXiv e-prints}\ ,\ \bibinfo {eid} {arXiv:2510.02247}} (\bibinfo {year} {2025})},\ \Eprint {https://arxiv.org/abs/2510.02247} {arXiv:2510.02247 [astro-ph.GA]} \BibitemShut {NoStop}%
\bibitem [{\citenamefont {{Gilman}}\ \emph {et~al.}(2021)\citenamefont {{Gilman}}, \citenamefont {{Bovy}}, \citenamefont {{Treu}}, \citenamefont {{Nierenberg}}, \citenamefont {{Birrer}}, \citenamefont {{Benson}},\ and\ \citenamefont {{Sameie}}}]{Gilman+21}%
  \BibitemOpen
  \bibfield  {author} {\bibinfo {author} {\bibfnamefont {D.}~\bibnamefont {{Gilman}}}, \bibinfo {author} {\bibfnamefont {J.}~\bibnamefont {{Bovy}}}, \bibinfo {author} {\bibfnamefont {T.}~\bibnamefont {{Treu}}}, \bibinfo {author} {\bibfnamefont {A.}~\bibnamefont {{Nierenberg}}}, \bibinfo {author} {\bibfnamefont {S.}~\bibnamefont {{Birrer}}}, \bibinfo {author} {\bibfnamefont {A.}~\bibnamefont {{Benson}}},\ and\ \bibinfo {author} {\bibfnamefont {O.}~\bibnamefont {{Sameie}}},\ }\bibfield  {title} {\bibinfo {title} {{Strong lensing signatures of self-interacting dark matter in low-mass haloes}},\ }\href {https://doi.org/10.1093/mnras/stab2335} {\bibfield  {journal} {\bibinfo  {journal} {\mnras}\ }\textbf {\bibinfo {volume} {507}},\ \bibinfo {pages} {2432} (\bibinfo {year} {2021})},\ \Eprint {https://arxiv.org/abs/2105.05259} {arXiv:2105.05259 [astro-ph.CO]} \BibitemShut {NoStop}%
\bibitem [{\citenamefont {{Gilman}}\ \emph {et~al.}(2023)\citenamefont {{Gilman}}, \citenamefont {{Zhong}},\ and\ \citenamefont {{Bovy}}}]{Gilman+23}%
  \BibitemOpen
  \bibfield  {author} {\bibinfo {author} {\bibfnamefont {D.}~\bibnamefont {{Gilman}}}, \bibinfo {author} {\bibfnamefont {Y.-M.}\ \bibnamefont {{Zhong}}},\ and\ \bibinfo {author} {\bibfnamefont {J.}~\bibnamefont {{Bovy}}},\ }\bibfield  {title} {\bibinfo {title} {{Constraining resonant dark matter self-interactions with strong gravitational lenses}},\ }\href {https://doi.org/10.1103/PhysRevD.107.103008} {\bibfield  {journal} {\bibinfo  {journal} {\prd}\ }\textbf {\bibinfo {volume} {107}},\ \bibinfo {eid} {103008} (\bibinfo {year} {2023})},\ \Eprint {https://arxiv.org/abs/2207.13111} {arXiv:2207.13111 [astro-ph.CO]} \BibitemShut {NoStop}%
\bibitem [{\citenamefont {{Enzi}}\ \emph {et~al.}(2025)\citenamefont {{Enzi}}, \citenamefont {{Krawczyk}}, \citenamefont {{Ballard}},\ and\ \citenamefont {{Collett}}}]{Enzi++25}%
  \BibitemOpen
  \bibfield  {author} {\bibinfo {author} {\bibfnamefont {W.~J.~R.}\ \bibnamefont {{Enzi}}}, \bibinfo {author} {\bibfnamefont {C.~M.}\ \bibnamefont {{Krawczyk}}}, \bibinfo {author} {\bibfnamefont {D.~J.}\ \bibnamefont {{Ballard}}},\ and\ \bibinfo {author} {\bibfnamefont {T.~E.}\ \bibnamefont {{Collett}}},\ }\bibfield  {title} {\bibinfo {title} {{The overconcentrated dark halo in the strong lens SDSS J0946 + 1006 is a subhalo: evidence for self-interacting dark matter?}},\ }\href {https://doi.org/10.1093/mnras/staf697} {\bibfield  {journal} {\bibinfo  {journal} {\mnras}\ }\textbf {\bibinfo {volume} {540}},\ \bibinfo {pages} {247} (\bibinfo {year} {2025})},\ \Eprint {https://arxiv.org/abs/2411.08565} {arXiv:2411.08565 [astro-ph.CO]} \BibitemShut {NoStop}%
\bibitem [{\citenamefont {{Powell}}\ \emph {et~al.}(2025)\citenamefont {{Powell}}, \citenamefont {{McKean}}, \citenamefont {{Vegetti}}, \citenamefont {{Spingola}}, \citenamefont {{White}},\ and\ \citenamefont {{Fassnacht}}}]{Powell++25}%
  \BibitemOpen
  \bibfield  {author} {\bibinfo {author} {\bibfnamefont {D.~M.}\ \bibnamefont {{Powell}}}, \bibinfo {author} {\bibfnamefont {J.~P.}\ \bibnamefont {{McKean}}}, \bibinfo {author} {\bibfnamefont {S.}~\bibnamefont {{Vegetti}}}, \bibinfo {author} {\bibfnamefont {C.}~\bibnamefont {{Spingola}}}, \bibinfo {author} {\bibfnamefont {S.~D.~M.}\ \bibnamefont {{White}}},\ and\ \bibinfo {author} {\bibfnamefont {C.~D.}\ \bibnamefont {{Fassnacht}}},\ }\bibfield  {title} {\bibinfo {title} {{A million-solar-mass object detected at a cosmological distance using gravitational imaging}},\ }\bibfield  {journal} {\bibinfo  {journal} {Nature Astronomy}\ }\href {https://doi.org/10.1038/s41550-025-02651-2} {10.1038/s41550-025-02651-2} (\bibinfo {year} {2025}),\ \Eprint {https://arxiv.org/abs/2510.07382} {arXiv:2510.07382 [astro-ph.CO]} \BibitemShut {NoStop}%
\bibitem [{\citenamefont {{Vogelsberger}}\ \emph {et~al.}(2016)\citenamefont {{Vogelsberger}}, \citenamefont {{Zavala}}, \citenamefont {{Cyr-Racine}}, \citenamefont {{Pfrommer}}, \citenamefont {{Bringmann}},\ and\ \citenamefont {{Sigurdson}}}]{Vogelsberger16}%
  \BibitemOpen
  \bibfield  {author} {\bibinfo {author} {\bibfnamefont {M.}~\bibnamefont {{Vogelsberger}}}, \bibinfo {author} {\bibfnamefont {J.}~\bibnamefont {{Zavala}}}, \bibinfo {author} {\bibfnamefont {F.-Y.}\ \bibnamefont {{Cyr-Racine}}}, \bibinfo {author} {\bibfnamefont {C.}~\bibnamefont {{Pfrommer}}}, \bibinfo {author} {\bibfnamefont {T.}~\bibnamefont {{Bringmann}}},\ and\ \bibinfo {author} {\bibfnamefont {K.}~\bibnamefont {{Sigurdson}}},\ }\bibfield  {title} {\bibinfo {title} {{ETHOS - an effective theory of structure formation: dark matter physics as a possible explanation of the small-scale CDM problems}},\ }\href {https://doi.org/10.1093/mnras/stw1076} {\bibfield  {journal} {\bibinfo  {journal} {\mnras}\ }\textbf {\bibinfo {volume} {460}},\ \bibinfo {pages} {1399} (\bibinfo {year} {2016})},\ \Eprint {https://arxiv.org/abs/1512.05349} {arXiv:1512.05349 [astro-ph.CO]} \BibitemShut {NoStop}%
\bibitem [{\citenamefont {{Nadler}}\ \emph {et~al.}(2024)\citenamefont {{Nadler}}, \citenamefont {{An}}, \citenamefont {{Yang}}, \citenamefont {{Yu}}, \citenamefont {{Benson}},\ and\ \citenamefont {{Gluscevic}}}]{Nadler++24}%
  \BibitemOpen
  \bibfield  {author} {\bibinfo {author} {\bibfnamefont {E.~O.}\ \bibnamefont {{Nadler}}}, \bibinfo {author} {\bibfnamefont {R.}~\bibnamefont {{An}}}, \bibinfo {author} {\bibfnamefont {D.}~\bibnamefont {{Yang}}}, \bibinfo {author} {\bibfnamefont {H.-B.}\ \bibnamefont {{Yu}}}, \bibinfo {author} {\bibfnamefont {A.}~\bibnamefont {{Benson}}},\ and\ \bibinfo {author} {\bibfnamefont {V.}~\bibnamefont {{Gluscevic}}},\ }\bibfield  {title} {\bibinfo {title} {{COZMIC. III. Cosmological Zoom-in Simulations of SIDM with Suppressed Initial Conditions}},\ }\href {https://doi.org/10.48550/arXiv.2412.13065} {\bibfield  {journal} {\bibinfo  {journal} {arXiv e-prints}\ ,\ \bibinfo {eid} {arXiv:2412.13065}} (\bibinfo {year} {2024})},\ \Eprint {https://arxiv.org/abs/2412.13065} {arXiv:2412.13065 [astro-ph.CO]} \BibitemShut {NoStop}%
\bibitem [{\citenamefont {{Tran}}\ \emph {et~al.}(2024{\natexlab{a}})\citenamefont {{Tran}}, \citenamefont {{Gilman}}, \citenamefont {{Vogelsberger}}, \citenamefont {{Shen}}, \citenamefont {{O'Neil}},\ and\ \citenamefont {{Zhang}}}]{Tran2024_1}%
  \BibitemOpen
  \bibfield  {author} {\bibinfo {author} {\bibfnamefont {V.}~\bibnamefont {{Tran}}}, \bibinfo {author} {\bibfnamefont {D.}~\bibnamefont {{Gilman}}}, \bibinfo {author} {\bibfnamefont {M.}~\bibnamefont {{Vogelsberger}}}, \bibinfo {author} {\bibfnamefont {X.}~\bibnamefont {{Shen}}}, \bibinfo {author} {\bibfnamefont {S.}~\bibnamefont {{O'Neil}}},\ and\ \bibinfo {author} {\bibfnamefont {X.}~\bibnamefont {{Zhang}}},\ }\bibfield  {title} {\bibinfo {title} {{Gravothermal catastrophe in resonant self-interacting dark matter models}},\ }\href {https://doi.org/10.1103/PhysRevD.110.043048} {\bibfield  {journal} {\bibinfo  {journal} {\prd}\ }\textbf {\bibinfo {volume} {110}},\ \bibinfo {eid} {043048} (\bibinfo {year} {2024}{\natexlab{a}})},\ \Eprint {https://arxiv.org/abs/2405.02388} {arXiv:2405.02388 [astro-ph.GA]} \BibitemShut {NoStop}%
\bibitem [{\citenamefont {{Tran}}\ \emph {et~al.}(2025)\citenamefont {{Tran}}, \citenamefont {{Shen}}, \citenamefont {{Gilman}}, \citenamefont {{Vogelsberger}}, \citenamefont {{O'Neil}}, \citenamefont {{Xiong}}, \citenamefont {{Hu}},\ and\ \citenamefont {{Wu}}}]{Tran2025_2}%
  \BibitemOpen
  \bibfield  {author} {\bibinfo {author} {\bibfnamefont {V.}~\bibnamefont {{Tran}}}, \bibinfo {author} {\bibfnamefont {X.}~\bibnamefont {{Shen}}}, \bibinfo {author} {\bibfnamefont {D.}~\bibnamefont {{Gilman}}}, \bibinfo {author} {\bibfnamefont {M.}~\bibnamefont {{Vogelsberger}}}, \bibinfo {author} {\bibfnamefont {S.}~\bibnamefont {{O'Neil}}}, \bibinfo {author} {\bibfnamefont {D.}~\bibnamefont {{Xiong}}}, \bibinfo {author} {\bibfnamefont {J.}~\bibnamefont {{Hu}}},\ and\ \bibinfo {author} {\bibfnamefont {Z.}~\bibnamefont {{Wu}}},\ }\bibfield  {title} {\bibinfo {title} {{Core collapse in resonant self-interacting dark matter across two decades in halo mass}},\ }\href {https://doi.org/10.1103/2p8b-qcys} {\bibfield  {journal} {\bibinfo  {journal} {\prd}\ }\textbf {\bibinfo {volume} {112}},\ \bibinfo {eid} {083003} (\bibinfo {year} {2025})},\ \Eprint {https://arxiv.org/abs/2504.02928} {arXiv:2504.02928 [astro-ph.GA]} \BibitemShut {NoStop}%
\bibitem [{\citenamefont {Yang}\ and\ \citenamefont {Yu}(2022)}]{Yang+2022}%
  \BibitemOpen
  \bibfield  {author} {\bibinfo {author} {\bibfnamefont {D.}~\bibnamefont {Yang}}\ and\ \bibinfo {author} {\bibfnamefont {H.-B.}\ \bibnamefont {Yu}},\ }\bibfield  {title} {\bibinfo {title} {Gravothermal evolution of dark matter halos with differential elastic scattering},\ }\href {https://doi.org/10.1088/1475-7516/2022/09/077} {\bibfield  {journal} {\bibinfo  {journal} {Journal of Cosmology and Astroparticle Physics}\ }\textbf {\bibinfo {volume} {2022}}\bibinfo  {number} { (09)},\ \bibinfo {pages} {077}}\BibitemShut {NoStop}%
\bibitem [{\citenamefont {{Yang}}\ \emph {et~al.}(2023)\citenamefont {{Yang}}, \citenamefont {{Du}}, \citenamefont {{Zeng}}, \citenamefont {{Benson}}, \citenamefont {{Jiang}}, \citenamefont {{Nadler}},\ and\ \citenamefont {{Peter}}}]{Yang+23}%
  \BibitemOpen
\bibfield  {number} {  }\bibfield  {author} {\bibinfo {author} {\bibfnamefont {S.}~\bibnamefont {{Yang}}}, \bibinfo {author} {\bibfnamefont {X.}~\bibnamefont {{Du}}}, \bibinfo {author} {\bibfnamefont {Z.~C.}\ \bibnamefont {{Zeng}}}, \bibinfo {author} {\bibfnamefont {A.}~\bibnamefont {{Benson}}}, \bibinfo {author} {\bibfnamefont {F.}~\bibnamefont {{Jiang}}}, \bibinfo {author} {\bibfnamefont {E.~O.}\ \bibnamefont {{Nadler}}},\ and\ \bibinfo {author} {\bibfnamefont {A.~H.~G.}\ \bibnamefont {{Peter}}},\ }\bibfield  {title} {\bibinfo {title} {{Gravothermal Solutions of SIDM Halos: Mapping from Constant to Velocity-dependent Cross Section}},\ }\href {https://doi.org/10.3847/1538-4357/acbd49} {\bibfield  {journal} {\bibinfo  {journal} {\apj}\ }\textbf {\bibinfo {volume} {946}},\ \bibinfo {eid} {47} (\bibinfo {year} {2023})},\ \Eprint {https://arxiv.org/abs/2205.02957} {arXiv:2205.02957 [astro-ph.CO]} \BibitemShut {NoStop}%
\bibitem [{\citenamefont {{Essig}}\ \emph {et~al.}(2019)\citenamefont {{Essig}}, \citenamefont {{McDermott}}, \citenamefont {{Yu}},\ and\ \citenamefont {{Zhong}}}]{Essig2019}%
  \BibitemOpen
  \bibfield  {author} {\bibinfo {author} {\bibfnamefont {R.}~\bibnamefont {{Essig}}}, \bibinfo {author} {\bibfnamefont {S.~D.}\ \bibnamefont {{McDermott}}}, \bibinfo {author} {\bibfnamefont {H.-B.}\ \bibnamefont {{Yu}}},\ and\ \bibinfo {author} {\bibfnamefont {Y.-M.}\ \bibnamefont {{Zhong}}},\ }\bibfield  {title} {\bibinfo {title} {{Constraining Dissipative Dark Matter Self-Interactions}},\ }\href {https://doi.org/10.1103/PhysRevLett.123.121102} {\bibfield  {journal} {\bibinfo  {journal} {\prl}\ }\textbf {\bibinfo {volume} {123}},\ \bibinfo {eid} {121102} (\bibinfo {year} {2019})},\ \Eprint {https://arxiv.org/abs/1809.01144} {arXiv:1809.01144 [hep-ph]} \BibitemShut {NoStop}%
\bibitem [{\citenamefont {{Yang}}\ \emph {et~al.}(2024)\citenamefont {{Yang}}, \citenamefont {{Nadler}}, \citenamefont {{Yu}},\ and\ \citenamefont {{Zhong}}}]{Yang2024}%
  \BibitemOpen
  \bibfield  {author} {\bibinfo {author} {\bibfnamefont {D.}~\bibnamefont {{Yang}}}, \bibinfo {author} {\bibfnamefont {E.~O.}\ \bibnamefont {{Nadler}}}, \bibinfo {author} {\bibfnamefont {H.-B.}\ \bibnamefont {{Yu}}},\ and\ \bibinfo {author} {\bibfnamefont {Y.-M.}\ \bibnamefont {{Zhong}}},\ }\bibfield  {title} {\bibinfo {title} {{A parametric model for self-interacting dark matter halos}},\ }\href {https://doi.org/10.1088/1475-7516/2024/02/032} {\bibfield  {journal} {\bibinfo  {journal} {\jcap}\ }\textbf {\bibinfo {volume} {2024}},\ \bibinfo {eid} {032} (\bibinfo {year} {2024})},\ \Eprint {https://arxiv.org/abs/2305.16176} {arXiv:2305.16176 [astro-ph.CO]} \BibitemShut {NoStop}%
\bibitem [{\citenamefont {{Tran}}\ \emph {et~al.}(2024{\natexlab{b}})\citenamefont {{Tran}}, \citenamefont {{Shen}}, \citenamefont {{Vogelsberger}}, \citenamefont {{Gilman}}, \citenamefont {{O'Neil}},\ and\ \citenamefont {{Gao}}}]{Tran2025_1}%
  \BibitemOpen
  \bibfield  {author} {\bibinfo {author} {\bibfnamefont {V.}~\bibnamefont {{Tran}}}, \bibinfo {author} {\bibfnamefont {X.}~\bibnamefont {{Shen}}}, \bibinfo {author} {\bibfnamefont {M.}~\bibnamefont {{Vogelsberger}}}, \bibinfo {author} {\bibfnamefont {D.}~\bibnamefont {{Gilman}}}, \bibinfo {author} {\bibfnamefont {S.}~\bibnamefont {{O'Neil}}},\ and\ \bibinfo {author} {\bibfnamefont {J.}~\bibnamefont {{Gao}}},\ }\bibfield  {title} {\bibinfo {title} {{A Novel Density Profile for Isothermal Cores of Dark Matter Halos}},\ }\href {https://doi.org/10.48550/arXiv.2411.11945} {\bibfield  {journal} {\bibinfo  {journal} {arXiv e-prints}\ ,\ \bibinfo {eid} {arXiv:2411.11945}} (\bibinfo {year} {2024}{\natexlab{b}})},\ \Eprint {https://arxiv.org/abs/2411.11945} {arXiv:2411.11945 [astro-ph.CO]} \BibitemShut {NoStop}%
\bibitem [{\citenamefont {{Springel}}(2010)}]{Springel2010}%
  \BibitemOpen
  \bibfield  {author} {\bibinfo {author} {\bibfnamefont {V.}~\bibnamefont {{Springel}}},\ }\bibfield  {title} {\bibinfo {title} {{E pur si muove: Galilean-invariant cosmological hydrodynamical simulations on a moving mesh}},\ }\href {https://doi.org/10.1111/j.1365-2966.2009.15715.x} {\bibfield  {journal} {\bibinfo  {journal} {\mnras}\ }\textbf {\bibinfo {volume} {401}},\ \bibinfo {pages} {791} (\bibinfo {year} {2010})},\ \Eprint {https://arxiv.org/abs/0901.4107} {arXiv:0901.4107 [astro-ph.CO]} \BibitemShut {NoStop}%
\bibitem [{\citenamefont {{Weinberger}}\ \emph {et~al.}(2020)\citenamefont {{Weinberger}}, \citenamefont {{Springel}},\ and\ \citenamefont {{Pakmor}}}]{Weinberger2020}%
  \BibitemOpen
  \bibfield  {author} {\bibinfo {author} {\bibfnamefont {R.}~\bibnamefont {{Weinberger}}}, \bibinfo {author} {\bibfnamefont {V.}~\bibnamefont {{Springel}}},\ and\ \bibinfo {author} {\bibfnamefont {R.}~\bibnamefont {{Pakmor}}},\ }\bibfield  {title} {\bibinfo {title} {{The AREPO Public Code Release}},\ }\href {https://doi.org/10.3847/1538-4365/ab908c} {\bibfield  {journal} {\bibinfo  {journal} {\apjs}\ }\textbf {\bibinfo {volume} {248}},\ \bibinfo {eid} {32} (\bibinfo {year} {2020})},\ \Eprint {https://arxiv.org/abs/1909.04667} {arXiv:1909.04667 [astro-ph.IM]} \BibitemShut {NoStop}%
\bibitem [{\citenamefont {{Mace}}\ \emph {et~al.}(2024)\citenamefont {{Mace}}, \citenamefont {{Carton Zeng}}, \citenamefont {{Peter}}, \citenamefont {{Du}}, \citenamefont {{Yang}}, \citenamefont {{Benson}},\ and\ \citenamefont {{Vogelsberger}}}]{Mace+24}%
  \BibitemOpen
  \bibfield  {author} {\bibinfo {author} {\bibfnamefont {C.}~\bibnamefont {{Mace}}}, \bibinfo {author} {\bibfnamefont {Z.}~\bibnamefont {{Carton Zeng}}}, \bibinfo {author} {\bibfnamefont {A.~H.~G.}\ \bibnamefont {{Peter}}}, \bibinfo {author} {\bibfnamefont {X.}~\bibnamefont {{Du}}}, \bibinfo {author} {\bibfnamefont {S.}~\bibnamefont {{Yang}}}, \bibinfo {author} {\bibfnamefont {A.}~\bibnamefont {{Benson}}},\ and\ \bibinfo {author} {\bibfnamefont {M.}~\bibnamefont {{Vogelsberger}}},\ }\bibfield  {title} {\bibinfo {title} {{Convergence Tests of Self-Interacting Dark Matter Simulations}},\ }\href {https://doi.org/10.48550/arXiv.2402.01604} {\bibfield  {journal} {\bibinfo  {journal} {arXiv e-prints}\ ,\ \bibinfo {eid} {arXiv:2402.01604}} (\bibinfo {year} {2024})},\ \Eprint {https://arxiv.org/abs/2402.01604} {arXiv:2402.01604 [astro-ph.GA]} \BibitemShut {NoStop}%
\bibitem [{\citenamefont {{Vogelsberger}}\ \emph {et~al.}(2012)\citenamefont {{Vogelsberger}}, \citenamefont {{Zavala}},\ and\ \citenamefont {{Loeb}}}]{Vogelsberger2012}%
  \BibitemOpen
  \bibfield  {author} {\bibinfo {author} {\bibfnamefont {M.}~\bibnamefont {{Vogelsberger}}}, \bibinfo {author} {\bibfnamefont {J.}~\bibnamefont {{Zavala}}},\ and\ \bibinfo {author} {\bibfnamefont {A.}~\bibnamefont {{Loeb}}},\ }\bibfield  {title} {\bibinfo {title} {{Subhaloes in self-interacting galactic dark matter haloes}},\ }\href {https://doi.org/10.1111/j.1365-2966.2012.21182.x} {\bibfield  {journal} {\bibinfo  {journal} {\mnras}\ }\textbf {\bibinfo {volume} {423}},\ \bibinfo {pages} {3740} (\bibinfo {year} {2012})},\ \Eprint {https://arxiv.org/abs/1201.5892} {arXiv:1201.5892 [astro-ph.CO]} \BibitemShut {NoStop}%
\bibitem [{\citenamefont {{Springel}}\ and\ \citenamefont {{White}}(1999)}]{Springel1999}%
  \BibitemOpen
  \bibfield  {author} {\bibinfo {author} {\bibfnamefont {V.}~\bibnamefont {{Springel}}}\ and\ \bibinfo {author} {\bibfnamefont {S.~D.~M.}\ \bibnamefont {{White}}},\ }\bibfield  {title} {\bibinfo {title} {{Tidal tails in cold dark matter cosmologies}},\ }\href {https://doi.org/10.1046/j.1365-8711.1999.02613.x} {\bibfield  {journal} {\bibinfo  {journal} {\mnras}\ }\textbf {\bibinfo {volume} {307}},\ \bibinfo {pages} {162} (\bibinfo {year} {1999})},\ \Eprint {https://arxiv.org/abs/astro-ph/9807320} {arXiv:astro-ph/9807320 [astro-ph]} \BibitemShut {NoStop}%
\bibitem [{\citenamefont {{Kazantzidis}}\ \emph {et~al.}(2004)\citenamefont {{Kazantzidis}}, \citenamefont {{Magorrian}},\ and\ \citenamefont {{Moore}}}]{Kazantzidis2004}%
  \BibitemOpen
  \bibfield  {author} {\bibinfo {author} {\bibfnamefont {S.}~\bibnamefont {{Kazantzidis}}}, \bibinfo {author} {\bibfnamefont {J.}~\bibnamefont {{Magorrian}}},\ and\ \bibinfo {author} {\bibfnamefont {B.}~\bibnamefont {{Moore}}},\ }\bibfield  {title} {\bibinfo {title} {{Generating Equilibrium Dark Matter Halos: Inadequacies of the Local Maxwellian Approximation}},\ }\href {https://doi.org/10.1086/380192} {\bibfield  {journal} {\bibinfo  {journal} {\apj}\ }\textbf {\bibinfo {volume} {601}},\ \bibinfo {pages} {37} (\bibinfo {year} {2004})},\ \Eprint {https://arxiv.org/abs/astro-ph/0309517} {arXiv:astro-ph/0309517 [astro-ph]} \BibitemShut {NoStop}%
\bibitem [{\citenamefont {{Lynden-Bell}}\ \emph {et~al.}(1968)\citenamefont {{Lynden-Bell}}, \citenamefont {Wood},\ and\ \citenamefont {Royal}}]{LyndenBell68}%
  \BibitemOpen
  \bibfield  {author} {\bibinfo {author} {\bibfnamefont {D.}~\bibnamefont {{Lynden-Bell}}}, \bibinfo {author} {\bibfnamefont {R.}~\bibnamefont {Wood}},\ and\ \bibinfo {author} {\bibfnamefont {A.}~\bibnamefont {Royal}},\ }\bibfield  {title} {\bibinfo {title} {The {{Gravo-Thermal Catastrophe}} in {{Isothermal Spheres}} and the {{Onset}} of {{Red-Giant Structure}} for {{Stellar Systems}}},\ }\href {https://doi.org/10.1093/mnras/138.4.495} {\bibfield  {journal} {\bibinfo  {journal} {Monthly Notices of the Royal Astronomical Society}\ }\textbf {\bibinfo {volume} {138}},\ \bibinfo {pages} {495} (\bibinfo {year} {1968})}\BibitemShut {NoStop}%
\bibitem [{\citenamefont {{Binney}}\ and\ \citenamefont {{Tremaine}}(1987)}]{Binney1987}%
  \BibitemOpen
  \bibfield  {author} {\bibinfo {author} {\bibfnamefont {J.}~\bibnamefont {{Binney}}}\ and\ \bibinfo {author} {\bibfnamefont {S.}~\bibnamefont {{Tremaine}}},\ }\href@noop {} {\emph {\bibinfo {title} {{Galactic dynamics}}}}\ (\bibinfo {year} {1987})\BibitemShut {NoStop}%
\bibitem [{\citenamefont {{Lacroix}}\ \emph {et~al.}(2018)\citenamefont {{Lacroix}}, \citenamefont {{Stref}},\ and\ \citenamefont {{Lavalle}}}]{Lacroix2018}%
  \BibitemOpen
  \bibfield  {author} {\bibinfo {author} {\bibfnamefont {T.}~\bibnamefont {{Lacroix}}}, \bibinfo {author} {\bibfnamefont {M.}~\bibnamefont {{Stref}}},\ and\ \bibinfo {author} {\bibfnamefont {J.}~\bibnamefont {{Lavalle}}},\ }\bibfield  {title} {\bibinfo {title} {{Anatomy of Eddington-like inversion methods in the context of dark matter searches}},\ }\href {https://doi.org/10.1088/1475-7516/2018/09/040} {\bibfield  {journal} {\bibinfo  {journal} {\jcap}\ }\textbf {\bibinfo {volume} {2018}},\ \bibinfo {eid} {040} (\bibinfo {year} {2018})},\ \Eprint {https://arxiv.org/abs/1805.02403} {arXiv:1805.02403 [astro-ph.GA]} \BibitemShut {NoStop}%
\bibitem [{\citenamefont {Aarseth}\ and\ \citenamefont {Hoyle}(1963)}]{Aarseth1963}%
  \BibitemOpen
  \bibfield  {author} {\bibinfo {author} {\bibfnamefont {S.~J.}\ \bibnamefont {Aarseth}}\ and\ \bibinfo {author} {\bibfnamefont {F.}~\bibnamefont {Hoyle}},\ }\bibfield  {title} {\bibinfo {title} {Dynamical evolution of clusters of galaxies, i},\ }\href {https://doi.org/10.1093/mnras/126.3.223} {\bibfield  {journal} {\bibinfo  {journal} {Monthly Notices of the Royal Astronomical Society}\ }\textbf {\bibinfo {volume} {126}},\ \bibinfo {pages} {223} (\bibinfo {year} {1963})},\ \Eprint {https://arxiv.org/abs/https://academic.oup.com/mnras/article-pdf/126/3/223/8074955/mnras126-0223.pdf} {https://academic.oup.com/mnras/article-pdf/126/3/223/8074955/mnras126-0223.pdf} \BibitemShut {NoStop}%
\bibitem [{\citenamefont {{Springel}}\ \emph {et~al.}(2001)\citenamefont {{Springel}}, \citenamefont {{Yoshida}},\ and\ \citenamefont {{White}}}]{Springel2001}%
  \BibitemOpen
  \bibfield  {author} {\bibinfo {author} {\bibfnamefont {V.}~\bibnamefont {{Springel}}}, \bibinfo {author} {\bibfnamefont {N.}~\bibnamefont {{Yoshida}}},\ and\ \bibinfo {author} {\bibfnamefont {S.~D.~M.}\ \bibnamefont {{White}}},\ }\bibfield  {title} {\bibinfo {title} {{GADGET: a code for collisionless and gasdynamical cosmological simulations}},\ }\href {https://doi.org/10.1016/S1384-1076(01)00042-2} {\bibfield  {journal} {\bibinfo  {journal} {\na}\ }\textbf {\bibinfo {volume} {6}},\ \bibinfo {pages} {79} (\bibinfo {year} {2001})},\ \Eprint {https://arxiv.org/abs/astro-ph/0003162} {arXiv:astro-ph/0003162 [astro-ph]} \BibitemShut {NoStop}%
\bibitem [{\citenamefont {{Monaghan}}\ and\ \citenamefont {{Lattanzio}}(1985)}]{Monaghan1985}%
  \BibitemOpen
  \bibfield  {author} {\bibinfo {author} {\bibfnamefont {J.~J.}\ \bibnamefont {{Monaghan}}}\ and\ \bibinfo {author} {\bibfnamefont {J.~C.}\ \bibnamefont {{Lattanzio}}},\ }\bibfield  {title} {\bibinfo {title} {{A refined particle method for astrophysical problems}},\ }\href@noop {} {\bibfield  {journal} {\bibinfo  {journal} {\aap}\ }\textbf {\bibinfo {volume} {149}},\ \bibinfo {pages} {135} (\bibinfo {year} {1985})}\BibitemShut {NoStop}%
\bibitem [{\citenamefont {{Zeng}}\ \emph {et~al.}(2022)\citenamefont {{Zeng}}, \citenamefont {{Peter}}, \citenamefont {{Du}}, \citenamefont {{Benson}}, \citenamefont {{Kim}}, \citenamefont {{Jiang}}, \citenamefont {{Cyr-Racine}},\ and\ \citenamefont {{Vogelsberger}}}]{Zeng+22}%
  \BibitemOpen
  \bibfield  {author} {\bibinfo {author} {\bibfnamefont {Z.~C.}\ \bibnamefont {{Zeng}}}, \bibinfo {author} {\bibfnamefont {A.~H.~G.}\ \bibnamefont {{Peter}}}, \bibinfo {author} {\bibfnamefont {X.}~\bibnamefont {{Du}}}, \bibinfo {author} {\bibfnamefont {A.}~\bibnamefont {{Benson}}}, \bibinfo {author} {\bibfnamefont {S.}~\bibnamefont {{Kim}}}, \bibinfo {author} {\bibfnamefont {F.}~\bibnamefont {{Jiang}}}, \bibinfo {author} {\bibfnamefont {F.-Y.}\ \bibnamefont {{Cyr-Racine}}},\ and\ \bibinfo {author} {\bibfnamefont {M.}~\bibnamefont {{Vogelsberger}}},\ }\bibfield  {title} {\bibinfo {title} {{Core-collapse, evaporation, and tidal effects: the life story of a self-interacting dark matter subhalo}},\ }\href {https://doi.org/10.1093/mnras/stac1094} {\bibfield  {journal} {\bibinfo  {journal} {\mnras}\ }\textbf {\bibinfo {volume} {513}},\ \bibinfo {pages} {4845} (\bibinfo {year} {2022})},\ \Eprint {https://arxiv.org/abs/2110.00259} {arXiv:2110.00259 [astro-ph.CO]} \BibitemShut {NoStop}%
\end{thebibliography}%

\appendix

\section{Initial conditions sampling}
\label{apd:ICs}

\subsection{Eddington's distribution function}
\label{sapd:Eddington}

In order to achieve dynamical equilibrium in the initial conditions (ICs) with respect to the CDM model, we utilize Eddington's inversion formula~\cite{Binney1987}
\begin{equation}
    \label{eqn:eddington_inversion}
    \rho\left(r\right) = \int_0^{\sqrt{2 \psi (r)}} 4 \pi v^2 {\rm d}v\,f\left( \xi \right).
\end{equation}
Here, $\xi = \psi \left(r\right) - v^2/2$ is the binding energy per unit of mass, $\psi$ is the negative gravitational potential, and $f\left( \xi \right)$ is a phase-space distribution function. We assume that the system is bounded and isotropic, thus $f$ only depends on $\xi$, and $f\left(<0\right) = 0$. The Eddington distribution function satisfying Equation \ref{eqn:eddington_inversion} takes the form
\begin{equation}
    \label{eqn:eddington_df}
    f \left( \xi \right) = \frac{1}{\sqrt{8} \pi^2} \int_{0}^{\xi} {\rm d}\psi \left[ \frac{1}{\sqrt{\xi-\psi}}\frac{{\rm d}^2\rho}{{\rm d}\psi^2} + \frac{1}{\sqrt{\xi}}\frac{{\rm d}\rho}{{\rm d}\psi} \Bigr\rvert_{\psi=0} \right].
\end{equation}
This can be evaluated from the analytical forms of $\rho\left(r\right)$ and $\psi\left(r\right)$ with $r$ as an intermediate variable. Integrating, we obtain the semi-analytical form of the Eddington distribution function. Similar to in~\citep{Tran2024_1}, we make use of a combination of analytical formulas and numerical calculations with cubic spline interpolations for the calculation of the Eddington distribution function. For the values of ${\rm d}^2\rho / {\rm d}\psi^2$, we make improvements in the calculations by utilizing the intermediate interpolations of ${\rm d}^2 \log{\rho} / {\rm d} \log{\psi}^2$ and ${\rm d} \log{\rho} / {\rm d} \log{\psi}$, which reduce the fractional density profile reconstruction $\left| \rho_{\rm{E}} - \rho \right| / \rho$ to $\sim 10^{-5}$ everywhere. Here, $\rho_{\rm{E}}$ and $\rho$ are the density profiles obtained from the Eddington distribution following Equation \ref{eqn:eddington_inversion} and the desired sampling density profile, respectively. The particle position and velocity samplings are performed using the inversion sampling method, instead of the rejection sampling utilized in~\citep{Tran2024_1}.

\subsection{ICs in the context of softened gravity}
\label{sapd:softened}

\begin{figure}
    \centering
    \includegraphics[width= 0.49 \textwidth]{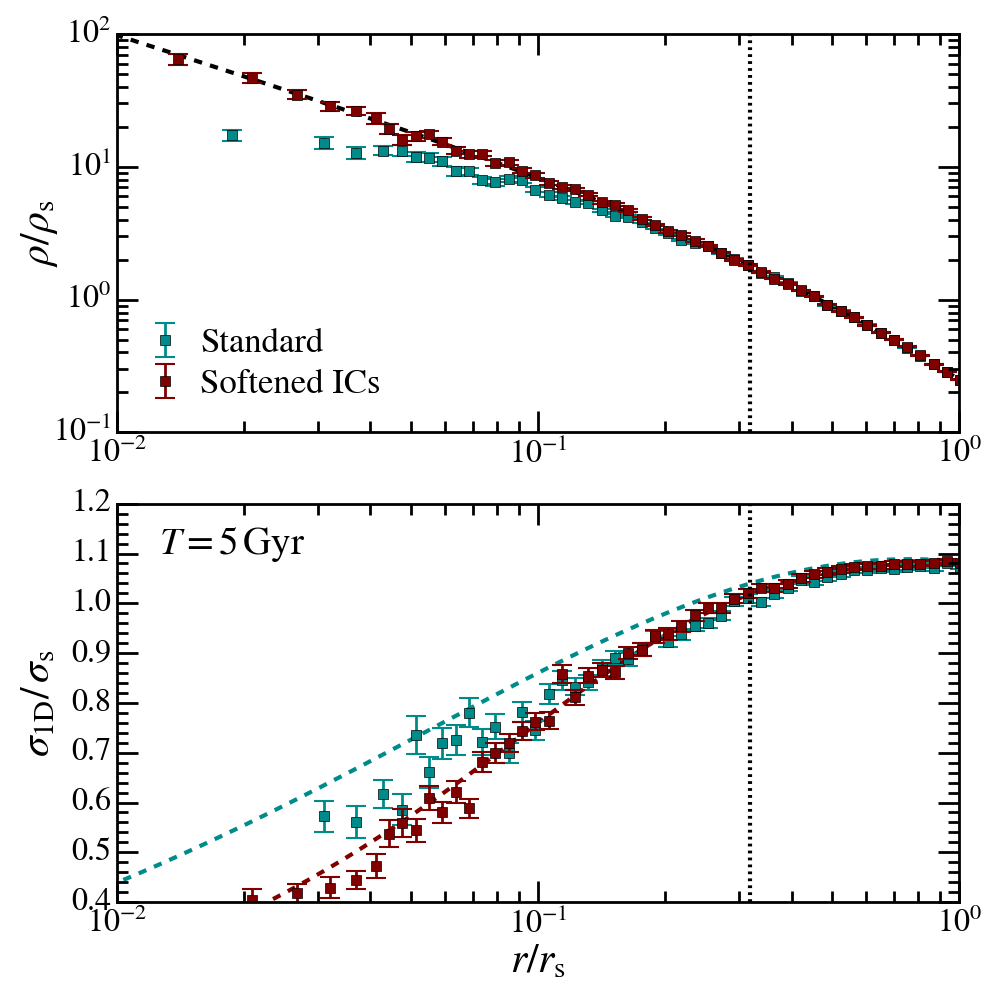}
    \caption{The density (top) and velocity dispersion (bottom) profiles of the SIDM-control halos initially sampled with $N_{200} = 10^6$. Results are shown for the standard Newtonian gravitational (blue) and the softened (pink) approaches. The halos are evolved for $5\Gyr$ in the CDM model with a Plummer-equivalent softening length of $\epsilon = 0.04\kpc$; the corresponding softening kernel scale, $h = 2.8 \epsilon$, is shown by the vertical dotted lines. Dashed lines denote the initial analytical profiles from which the particles are sampled. While the standard halo evolves toward a significantly different stable configuration from the initial profile, the softened halo retains its density and velocity dispersion profiles.}
    \label{fig:softened_core}
\end{figure}

Typically, the Eddington distributions are constructed with the Newtonian gravitational potential. However, under the softened gravity schemes employed in simulations, even without DM self-interaction, halos sampled using such distributions would become unstable, leading to the formation of small cores roughly the size of the softening kernel~\citep[e.g.][]{Tran2024_1,Lacroix2018}. This is shown in Figure \ref{fig:softened_core}, where we evolved the SIDM-control halo in the CDM model under a large softening length, $10$ times the recommended value in~\citep{Mace+24}. 

In order to appropriately account for this softening effect, we replace the Newtonian gravitational potential utilized in the calculation in the Eddington distribution function $\mathcal{f} (\xi)$ (Equation \ref{eqn:eddington_df}) with a softened gravitational field. The process is as follows. For any density kernel $W (x)$, such that a point mass $m$ is replaced by a spherical density profile $\rho(r) = m W(r)$, the convoluted density at the radius of $r$ resulting from a uniformly distributed spherical surface of mass $M$ and radius $R$ follows
\begin{equation}
    \label{eqn:convoluted_shell_init}
    \rho_\Sigma (r) = \frac{M}{4 \pi R^2} \int_0^\pi W (x(\theta)) \, 2 \pi R^2 \sin{\theta} {\rm{d}}\theta,
\end{equation}
where, $x (\theta)^2 = r^2 + R^2 - 2 r R \cos{\theta}$. Changing the integral variable from $\theta$ to $x$ using $\sin{\theta} {\rm{d}}\theta = x {\rm{d}}x / r R$, we have
\begin{equation}
    \label{eqn:convoluted_shell}
    \rho_\Sigma (r) = \frac{M}{2 r R} \int_{\left| r - R \right|}^{r + R} x W(x) {\rm{d}}x.
\end{equation}
For the Plummer model~\citep{Aarseth1963}, the kernel takes the form of
\begin{equation}
    \label{eqn:Plummer_kernel}
    W_{\rm{P}} (x; \epsilon) = \frac{3}{4 \pi \epsilon^3} \left( 1 + \frac{r^2}{\epsilon^2} \right)^{-5/2},
\end{equation}
which results in the commonly seen softened gravitational potential of $\Phi_{\rm{P}} (r) = - G m / \sqrt{r^2 + \epsilon^2}$. However, such a model converges relatively slowly to the Newtonian case of $\Phi (r) = Gm/r$, leading to codes such as AREPO~\citep{Weinberger2020} and GADGET~\citep{Springel2001} preferring a spline approach~\citep{Monaghan1985}, with $W_{\rm{S}} (x)$ taking the form of
\begin{equation}
    \label{eqn:Spline_kernel}
    W_{\rm{S}} (x; h) = \frac{8}{\pi h^3}
    \begin{cases}
        1 - 6 \left(\frac{x}{h}\right)^2 + 6 \left(\frac{x}{h}\right)^3, & 0 \leq \frac{x}{h} < \frac{1}{2} \\
        2 \left(1 - \frac{x}{h}\right)^3, & \frac{1}{2} \leq \frac{x}{h} < 1 \\
        0, & \frac{x}{h} \geq 1
    \end{cases}
\end{equation}
Applying this to Equation \ref{eqn:convoluted_shell}, we obtain
\begin{equation}
    \label{eqn:Spline_shell}
    \rho_{\Sigma,\rm{S}} (r) = \frac{4m}{\pi h r R} \mathcal{I} (u_1, u_2),
\end{equation}
with $u_1 = \left|r - R\right| / h$, $u_2 = \max{\left(1, \, \left(r + R\right) / h\right)}$, and
\begin{equation}
    \label{eqn:I}
    \mathcal{I} (u_1, u_2) = 
    \begin{cases}
        A(u_2) - A(u_1), & u_2 < \frac{1}{2} \\
        B(u_2) - B(\frac{1}{2}) + A(\frac{1}{2}) - A(u_1), & u_1 \leq \frac{1}{2} \leq u_2 \\
        B(u_2) - B(u_1), & u_1 \geq \frac{1}{2} \\
        0, & u_1 \geq u_2
    \end{cases}
\end{equation}
Here, the final case occurs when $r$ is outside of the kernel width, i.e., $\left| r - R \right| \geq h$. The two auxiliary equations $A (u)$ and $B (u)$ are defined as
\begin{align}
    \label{eqn:A}
    A (u) &= \frac{1}{2} u^2 - \frac{3}{2} u^4 + \frac{6}{5} u^5 , \\
    \label{eqn:B}
    B (u) &= u^2 - 2 u^3 + \frac{3}{2} u^4 - \frac{2}{5} u^5. 
\end{align}

In simulations that employ such a spline softening approach, such as those in AREPO and GADGET, the gravitational field resulting from an isotropic density profile $\rho (r)$ would be equivalent to the Newtonian field generated by the convoluted density profile
\begin{multline}
    \label{eqn:convoluted_halo}
    \rho_{\rm{S}} (r) = \int_0^\infty \frac{16 r^\prime \rho (r^\prime) \, {\rm{d}}r^\prime}{h r} \\ 
    \mathcal{I} \left( \left| r - r^\prime \right| / h, \, \max{\left(1, \, \left(r + r^\prime\right) / h\right)} \right).
\end{multline}
The gravitational potential would be calculated as
\begin{multline}
    \label{eqn:gravitational_potential}
    \Phi_{\rm{S}} (r) = -G \left[ \frac{1}{r} \int_{0}^{r} 4\pi {r^\prime}^2 \rho_{\rm{S}}(r^\prime) \, {\rm{d}} r^\prime + \right. \\
    \left. \int_{r}^{\infty} 4\pi r^\prime \rho_{\rm{S}}(r^\prime) \, {\rm{d}}r^\prime \right].
\end{multline}

Halos sampled using the Eddington distribution calculated with this softened gravitational potential remain stable, given that enough particles are enclosed within the softening kernel (such that the velocity structure is well sampled, and the two-body relaxation effect is avoided). This is again shown in Figure \ref{fig:softened_core}, with the softened ICs remaining consistent with the initially sampled configuration, while the standard Newtonian approach leads to halos reaching stable configurations which deviate significantly from the initial profiles.

\section{Accelerated evolution due to outer halo stripping}
\label{apd:stripping}

\begin{figure}
    \centering
    \includegraphics[width= 0.49 \textwidth]{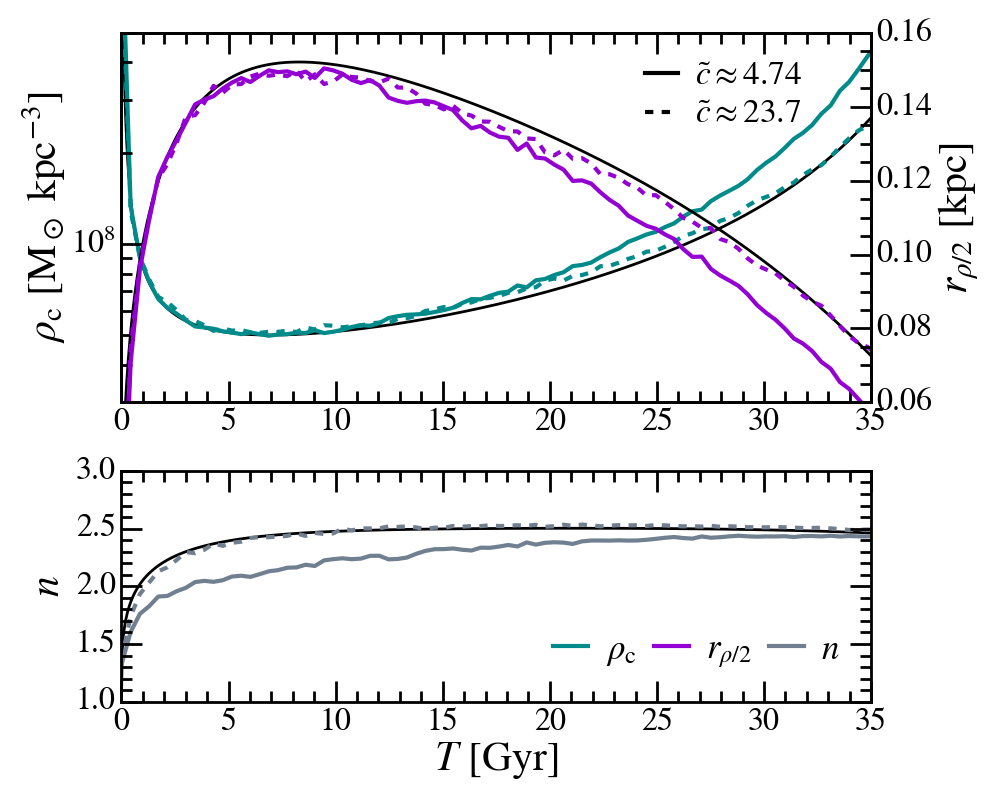}
    \caption{Comparison of the evolutions of the core-density $\rho_{\rm{c}}$ (top, cyan), core half-density radius $r_{\rho/2}$ (top, purple), and transition index $n$ (bottom, grey) between the SIDM-control halo with $\tilde{c} \approx 4.74$ (solid) and an extended version with $\tilde{c} \approx 23.7$ (dashed). The effective concentration $\tilde{c}$ is defined as the ratio between the NFW-exponential cut-off boundary and $r_{-2}$. The Plummer-equivalent softening length and timestep criteria are kept identical in both cases. The SIDM-control halo traces the same evolutionary track as the extended halo and the $\tau = 43.13\Gyr$ self-similar collapse track identified in~\citep{Tran2025_1} (black) during the core-formation phase, but exhibits an acceleration of approximately $10\%$ during the core-collapse phase. The transition index of the SIDM-control halo also attains lower values compared to the references.}
    \label{fig:extended_evolution}
\end{figure}

Figure \ref{fig:extended_evolution} compares the core evolution track of the SIDM-control halo in the VICS model with the approximate self-similar track from \citep{Tran2025_1}, obtained using the predicted timescale of $\tau = 43.13\Gyr$. For comparison, we also include an “extended” version of the SIDM-control halo, initialized with an NFW profile (Equation \ref{eqn:inner_density}) out to $5,r_{200}$, rather than switching to the exponential cutoff at $r_{200}$. This extended halo effectively represents a more concentrated configuration in the cosmological context. Initially, the core evolution of both halos follows the self-similar track closely. However, as the systems enter the core-collapse phase, deviations emerge. While the extended halo continues to follow the self-similar reference, the SIDM-control halo exhibits an approximately $10\%$ faster collapse. Moreover, we observe the transition index of the SIDM-control halo reaches lower values, indicating a relative deficiency of material in the outer regions, which weakens the contribution from the $\rho \propto r^{-3}$ regime. This deficit also results in a cooler outer region, promoting faster outward heat transport and thereby accelerating the gravothermal collapse. The fact that these deviations arise only in the later stages can be attributed to the varying collision timescales across the halo: at larger radii, where both the density and velocity dispersion are lower, collisions occur less frequently, delaying their influence on the evolution of the core. This is consistent with the effects of tidal stripping observed in cosmological simulations~\citep[e.g.][]{Zeng+22}.

\section{The effective NFW parameters of prompt cusp halos}
\label{apd:eff_params}

\begin{figure}
    \centering
    \includegraphics[width= 0.49 \textwidth]{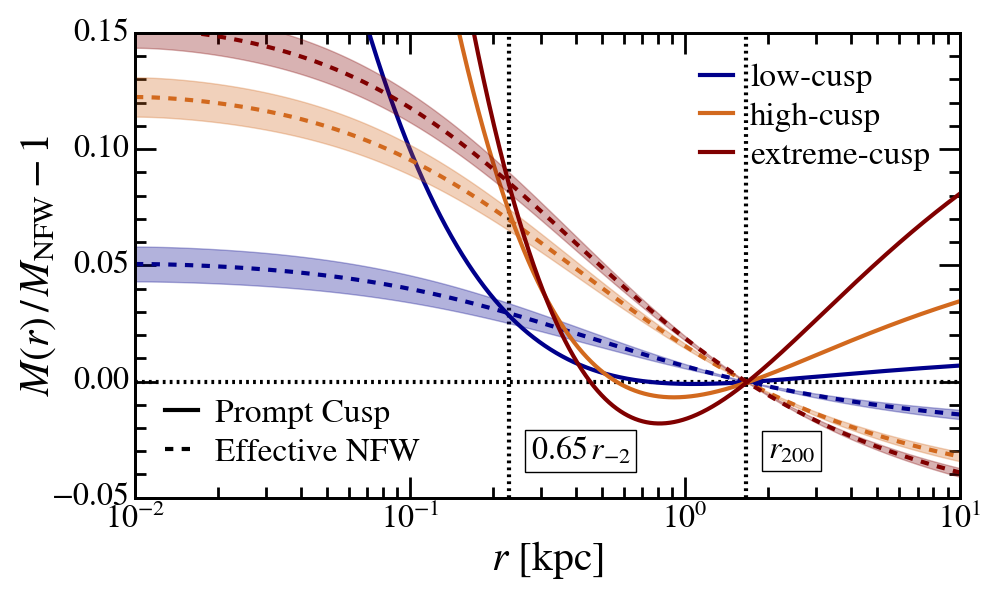}
    \caption{The fractional differences of the extreme-cusp (red), high-cusp (orange), and low-cusp (blue) initial mass profiles compared to the reference NFW halo (SIDM-control). The contracted effective NFW profiles of the same mass as our halos and concentration parameters $c = \bar{c}$ as detailed in Table \ref{tab:collapse_params} are shown with the dashed lines. The colored regions indicate the standard deviation resulting from uncertainties in $c$. The two vertical dotted lines show the virial radius $r_{200}$ (right) and the mass-matching radius of $0.65 \, r_{-2}$ (left).}
    \label{fig:eff_density_profiles}
\end{figure}

As detailed in Table \ref{tab:collapse_params}, the effective scale densities $\bar{\rho}{\rm{s}}$ and scale radii $\bar{r}{\rm{s}}$ approximately preserve the virial mass of the halo, even in the extreme-cusp case, and together correspond to an effective concentration. In order to predict such concentrations in future work, we examine the enclosed mass profiles of the effective NFW halos, constructed using the effective concentrations detailed in Table \ref{tab:collapse_params}, and the initial prompt cusp configurations. Figure \ref{fig:eff_density_profiles} presents the fractional differences of these profiles relative to the reference NFW profile of the SIDM-control halo. We observe that, across varying prompt cusp prominences, the enclosed masses of the prompt cusp profiles consistently coincide with those of the effective NFW halos at approximately $0.65 \, r_{-2}$. We therefore define this radius as the mass-matching radius $r_{\bar{c}}$ and adopt it for use in future analyses.

\section{Supplemental prompt-cusp halos evolution}
\label{apd:extra_halo}

\begin{figure}
    \centering
    \includegraphics[width= 0.49 \textwidth]{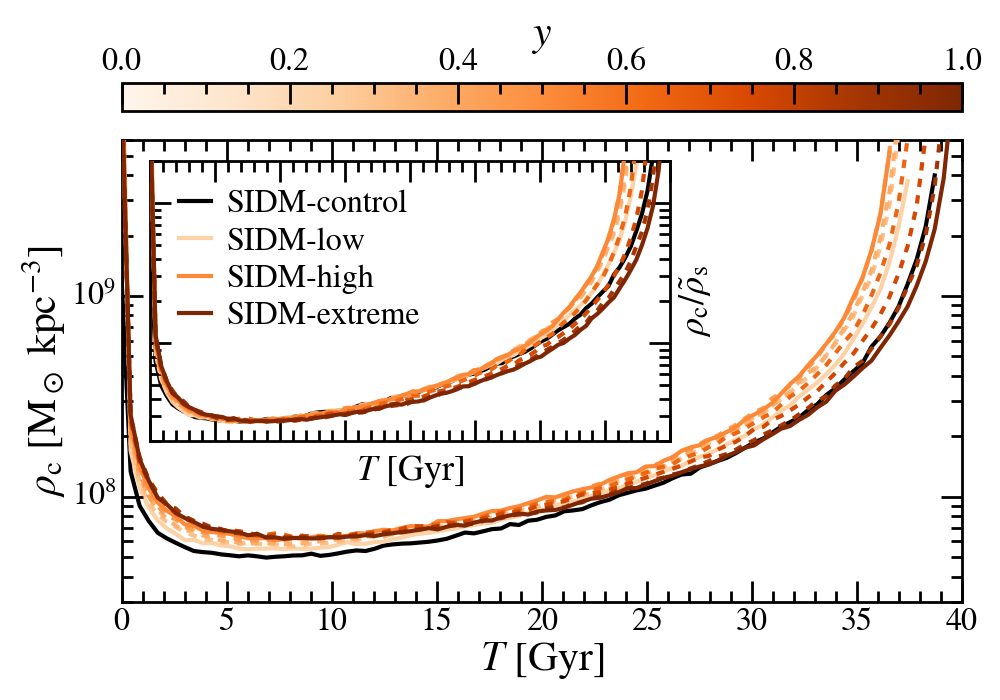}
    \caption{Core density evolution of prompt-cusp halos with varying levels of PC prominence, shown in physical units (main panel) and in terms of the effective scale density (sub-panel). As observed in Figures \ref{fig:core_evolution} and \ref{fig:core_evolution_scaled}, the collapse time depends non-trivially on both the effective concentration and the configuration of the outer regions. The smooth variation of the collapse time as a function of $y$ further indicates that this effect is physical rather than numerical.}
    \label{fig:extra_core_evolution}
\end{figure}

\begin{figure}
    \centering
    \includegraphics[width= 0.49 \textwidth]{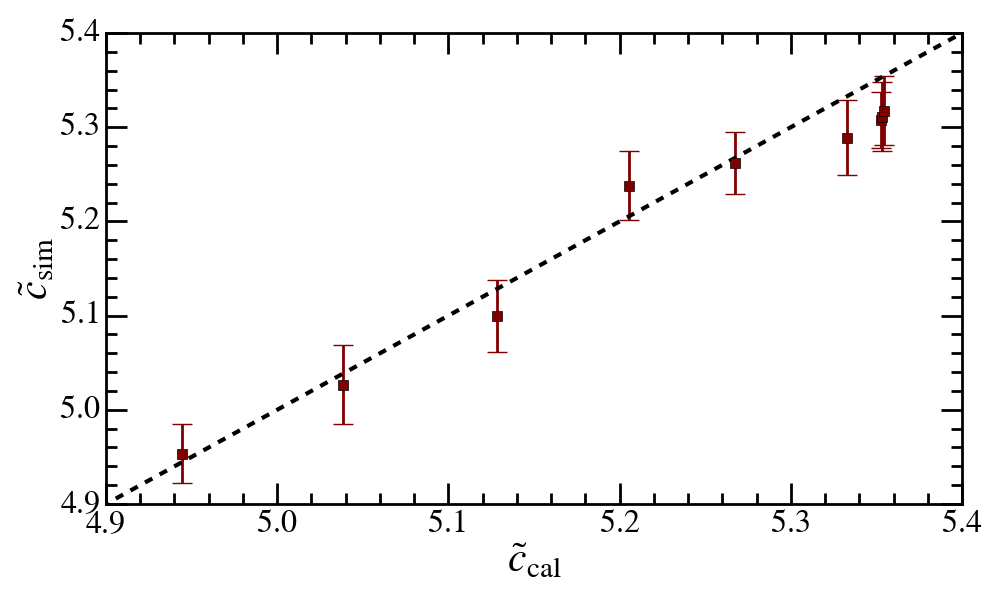}
    \caption{Comparison between the effective NFW concentrations calculated following Appendix \ref{apd:eff_params} and those measured following Section \ref{ssec:structure} for the halos presented in Figure \ref{fig:extra_core_evolution}. The dashed line indicates the case of perfect agreement, which is nearly achieved.}
    \label{fig:eff_c}
\end{figure}

\begin{table}
    \centering
    \addtolength{\tabcolsep}{24pt}
    \def\arraystretch{1.8}
    \begin{tabular}{c c c}
        \hline
        $r_{\rm{s}}$ & $\log \rho_{\rm{s}}$ & $y$ \\ [0ex] %
        [$\rm kpc$] & [${\rm M}_\odot\,{\rm kpc}^{-3}$] &  \\ [1ex] %
        \hline\hline

        0.387 & 7.186 & 0.310 \\
        
        0.405 & 7.133 & 0.378 \\
        
        0.427 & 7.072 & 0.445 \\

        0.511 & 6.867 & 0.635 \\
        
        0.576 & 6.726 & 0.756 \\
        
        0.643 & 6.595 & 0.878 \\

        \hline
    \end{tabular}
    \caption{Halo configurations for the additional prompt-cusp halos evolved in Appendix \ref{apd:extra_halo}. The numerical setups of these runs are similar to those adopted in the VICS run described in Section \ref{ssec:profilesetup}.}
    \label{tab:extra_runs_config}
\end{table}

In order to further support the points made in Section \ref{ssec:structure} and Appendix \ref{apd:eff_params}, we evolve six additional halos with configurations detailed in Table \ref{tab:extra_runs_config}. The parameters are chosen such that the PC prominence of these halos lies between those of the SIDM-low and SIDM-extreme cases. Figure \ref{fig:extra_core_evolution} shows the evolution of the core density for these halos, alongside the evolution of the halos presented in the main text. As expected from the dual influence of the increased concentration in the inner regions and the elevated temperature of the outer regions, the collapse time varies smoothly but non-trivially. This supports the argument made in Section \ref{ssec:core-collapse}. Figure \ref{fig:eff_c} shows the effective NFW concentrations measured directly from the simulations, as well as those calculated following the predictions of Appendix \ref{apd:eff_params}. These values are consistent and demonstrate that the mass within $\simeq 0.65 \, r_{-2}$ can be used to predict the behavior of halos with PC (and potentially other configurations) during the core formation phase.

\section{Halo comparison in the VDCS model}
\label{apd:vdcs_profiles}

\begin{figure}
    \centering
    \includegraphics[width= 0.49 \textwidth]{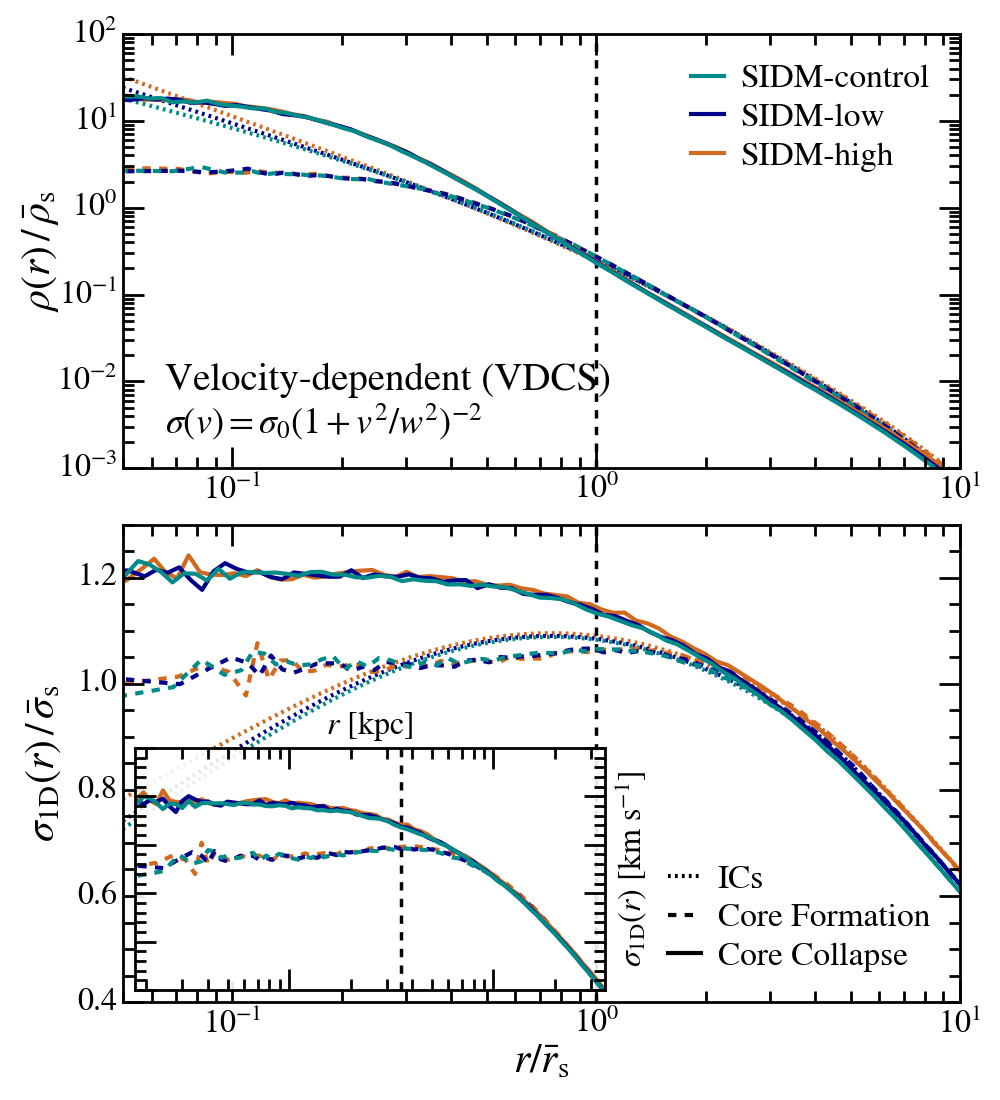}
    \caption{Similar to Figure \ref{fig:profiles}, but with halos in the VDCS model. The core-formation halos are chosen such that $\rho_{\rm{c}} / \bar{\rho}_{\rm{s}} \simeq 2.80$ and $r_{\rho/2} / \bar{r}_{\rm{s}} \simeq 0.387$, while the core-collapse halos satisfying $\rho_{\rm{c}} / \bar{\rho}_{\rm{s}} \simeq 20.0$ and $r_{\rho/2} / \bar{r}_{\rm{s}} \simeq 0.169$.}
    \label{fig:profiles_vdcs}
\end{figure}


Figure \ref{fig:profiles_vdcs} presents the density and one-dimensional velocity dispersion profiles of the control NFW halo, as well as the high- and low-cusp halos, in the VDCS model, taken at representative snapshots during the core-formation and core-collapse stages. The same conclusions as in the case of halo evolved with VICS can also be reached here.

\end{document}